%% file: main_UNANONYMIZED.tex
\documentclass[12pt]{iopart}
\usepackage{tikz}
\usepackage{graphicx}
\usepackage{url}
\usepackage{fancyvrb}
\usepackage{color, soul}
\usepackage{tikz}
\usepackage{hyperref}
\usepackage{float}
\usepackage{changepage}

\usetikzlibrary{shapes,arrows,positioning,shapes.geometric}
\usepackage{charter}
\usepackage{siunitx}
\usepackage{amssymb, bm}
\usepackage{graphicx}
\usepackage{color}
\usepackage{afterpage}
\usepackage{cleveref}
\usepackage{array,multirow}

\usepackage{struktex}

\usepackage{xspace}
\usepackage[colorinlistoftodos, prependcaption, textsize=tiny]{todonotes}

\usepackage{array}
\newcolumntype{P}[1]{>{\centering\arraybackslash}p{#1}}
\newcolumntype{M}[1]{>{\centering\arraybackslash}m{#1}}

\newcommand{\BE}[0]{\begin{equation}}
\newcommand{\EE}[0]{\end{equation}}
\newcommand{\BEA}[0]{\begin{eqnarray}}
\newcommand{\EEA}[0]{\end{eqnarray}}

\newcommand{\nuebar}{\ensuremath{\bar{\nu}_e}\xspace}

\mathchardef\mhyphen="2D

\newcommand{\orcid}[1]{\href{https://orcid.org/#1}{\includegraphics[height=\fontcharht\font`\B]{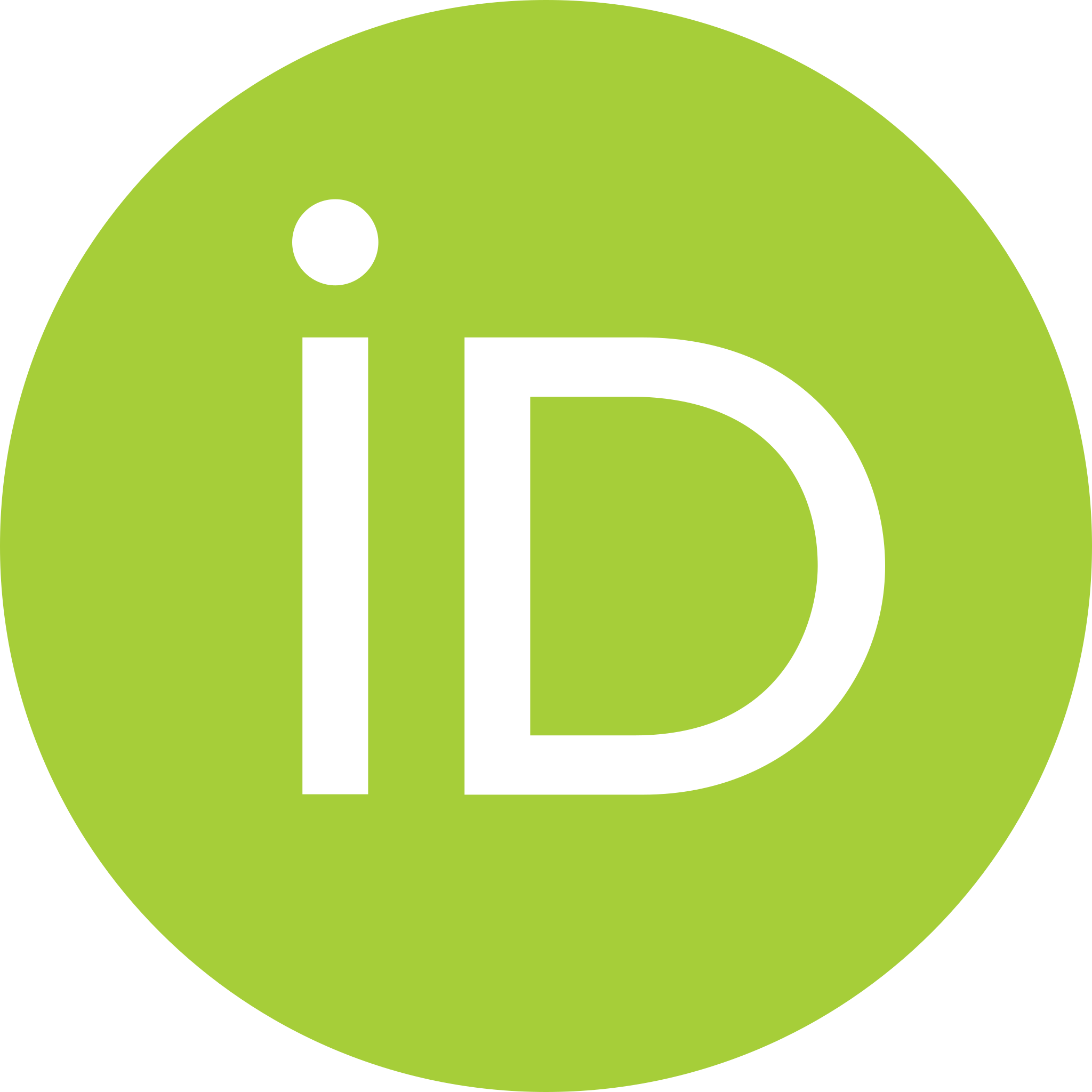}}}

\sethlcolor{red}

\begin{document}


\title[Surrogate Modeling of Radio-Frequency Quadrupole Particle Accelerators]
      {Neural Networks as Effective Surrogate Models of Radio-Frequency 
       Quadrupole Particle Accelerator Simulations}

\author{Joshua Villarreal$^1$ \orcid{0000-0001-9690-1310}, 
        Daniel Winklehner$^1$ \orcid{0000-0002-0715-6310}, Daniel Koser$^1$ \orcid{0009-0005-3062-0360}
        Janet M. Conrad$^1$ \orcid{0000-0002-6393-0438}}
\address{$^1$Massachusetts Institute of Technology, 77 Massachusetts Ave, 
             Cambridge, MA 02139, USA}
\ead{villaj@mit.edu}
\vspace{10pt}



\begin{abstract}
Radio-Frequency Quadrupoles (RFQs) are multi-purpose linear particle accelerators that simultaneously bunch and accelerate charged particle beams. They are ubiquitous in accelerator physics, especially as injectors to higher-energy machines, owing to their impressive efficiency. The design and optimization of these devices can be lengthy due to the need to repeatedly perform high-fidelity simulations. Several recent papers have demonstrated that machine learning can be used to build surrogate models (fast-executing replacements of computationally costly beam simulations) for order-of-magnitude computing time speedups. However, while these pilot studies are encouraging, there is room to improve their predictive accuracy. Particularly, beam summary statistics such as emittances (an important figure of merit in particle accelerator physics) have historically been challenging to predict. For the first time, we present a
 surrogate model trained on 200,000 samples that yields $<6\,\%$ mean average percent error for the predictions of all relevant beam output parameters from defining RFQ design parameters, solving the problem of poor emittance predictions by identifying and including hidden variables which were not accounted for previously. These surrogate models were made possible by using the Julia language and GPU computing; we briefly discuss both. We demonstrate the utility of surrogate modeling by performing a multi-objective optimization using our best model as a callback in the objective function to select an optimal RFQ design. We consider trade-offs in RFQ performance for various choices of Pareto-optimal design variables---common issues for any multi-objective optimization scheme. Lastly, we make recommendations for input data preparation, selection, and neural network architectures that pave the way for future development of production-capable surrogate models for RFQs and other particle accelerators.
\end{abstract}


\maketitle


\input{Sec1_Introduction}

\input{Sec2_Methods}

\input{Sec3_Results}

\input{Sec4_Discussion}

\input{Sec5_Conclusion}

\section*{Acknowledgments}
We would like to thank Andreas Adelmann for valuable input on 
the topic of surrogate modeling of particle accelerators.
We are grateful to the broader Julia community, whose 
expertise and responsiveness helped make this work possible.

This work was supported by NSF grants PHY-1505858 and 
PHY-1626069. DW was supported by funding from the Bose 
Foundation and the Heising-Simons Foundation.

\newpage

\section*{References}
\bibliographystyle{iopart-num}
\bibliography{Bibliography}

\appendix
\input{appendix}

\end{document}

%% file: Sec1_Introduction.tex
\section{Introduction}
\label{sec:intro}
The replacement of highly accurate, but computationally costly, particle-in-cell 
simulations with \emph{surrogate models} (sometimes called 
\emph{virtual accelerators}) is a topical field of increased interest~\cite{edelen:ml2,adelmann-2019-1,vay:icfa,sagan:icfa,bellotti:ml1,adelmann:snowmass21}.
Surrogate models use machine learning (ML)
to create fast-executing virtual representations of a complex system
like a particle accelerator.
We can then use this surrogate model to, for instance, speed up
(multi-objective) design optimization, or obtain real-time feedback
during the commissioning, tuning, and running of the particle 
accelerator. The surrogate model is typically built from a neural network (NN) or some other statistical learning technique (like polynomial chaos expansion~\cite{adelmann-2019-1}). In the case of using the surrogate model as an autonomous
tuning tool, training data can be obtained not just from simulations, but also by 
measurements from existing hardware~\cite{guptaImprovingSurrogateModel2021}.

The design of the IsoDAR (isotope decay-at-rest) project~\cite{bungau:isodar1, alonso:isodar_jinst, alonso:isodar_prd},
a planned experiment in neutrino physics, is the primary motivation for this work.
In IsoDAR, a compact particle accelerator produces a 10~mA proton 
beam that 
impinges with an energy of 60~MeV/amu on a beryllium target surrounded by lithium-7, producing electron antineutrinos (\nuebar) with a well-understood energy distribution through isotope decay-at-rest \cite{alonso:isodar_prd} (as opposed to
other experiments using decay-in-flight).
The \nuebar can then be measured in a nearby
liquid scintillator detector via inverse beta decay (IBD). This configuration yields unprecedented sensitivity to
so-called \emph{sterile neutrinos}, hypothesized new particles thought to resolve \nuebar deficits observed at experiments worldwide \cite{reactoranomaly, dayabay, microbooneanomaly, bestanomaly}.
The requirements for IsoDAR are 10~mA of protons on target in a continuous wave beam at 80\% duty factor to produce about $1.15\cdot10^{23}$~\nuebar over the course of 5 years. 
Paired with the planned 2.3~kiloton liquid scintillator detector (LSC)~\cite{seo_physics_2023}
in Korea, this will yield 1.67 million IBD events in the detector.

The IsoDAR particle accelerator comprises an ion source, radio-frequency quadrupole (RFQ), and a cyclotron \cite{winklehner:nima, winklehner:njp}. Surrogate modeling has proven invaluable to IsoDAR's development, allowing us to demonstrate the robustness of IsoDAR's cyclotron design~\cite{winklehner:njp} through uncertainty quantification~\cite{adelmann-2019-1}, and to perform a small pilot study to investigate the use of surrogate models for RFQs~\cite{frontiers}.

In this paper, we expand upon work presented in Ref.~\cite{frontiers} to build a neural network-based surrogate model of an RFQ, but rethink the RFQ parametrization to account for collinear effects in the feature space, physical RFQ design constraints, and incorporate variables previously hidden from trained surrogate models.
We use these insights to generate an accurate surrogate model for 
a 32.8~MHz RFQ covering a wide design parameter space (subject to physical design constraints).
We use the highly efficient Julia programming 
language~\cite{Julia-2017} to train our NNs with 
a widely cast net for hyperparameter tuning and an unusually large 
batch size.

The working principle of an RFQ and the generation of training 
data for the surrogate model were discussed in detail in 
Ref.~\cite{frontiers}. We briefly 
summarize them in Sec.~\ref{sec:rfq}, and Sec.~\ref{sec:datagen},
respectively. In Sec.~\ref{sec:methods}, we discuss our methods,
including data preparation and how we enhance the predictive accuracy of beam summary statistics
outputs like the beam emittance 
(an important figure of merit for beam quality), followed by our results 
for training of several NNs and optimization of the RFQ in 
Sec.~\ref{sec:results}. Finally, we discuss our results,
general observations, and recommendations for future development of NN-powered surrogate models in
Sec.~\ref{sec:discussion}-~\ref{sec:conclusion}. Code relevant to this project is available on
GitHub~\cite{rfqnn-github}.

\subsection{The Radio-Frequency Quadrupole}
\label{sec:rfq}

\begin{figure*}[t!]
    \begin{center}
        \includegraphics[width=0.6\columnwidth]
                        {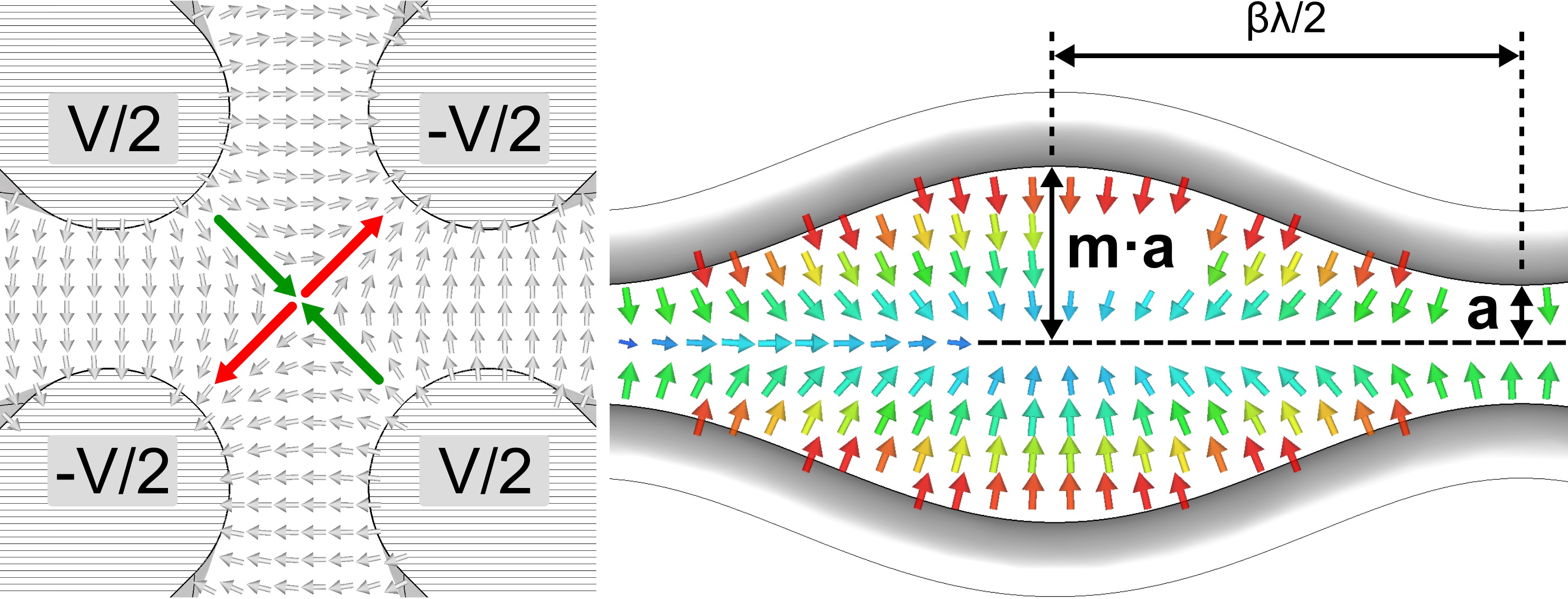}
        \caption{A single RFQ cell with electric fields in quasi-static approximation. Left: Front view (beam and $z$-axis point into the paper). The green arrows indicate the focusing of the charged particle beam and the red arrows the defocusing. Right: Side view (beam propagates from left to right). Cell
        geometry parameters are shown, and the electric field is indicated by arrows. From Ref.~\cite{frontiers}.}
        \label{fig:rfq_cell}
    \end{center}
\end{figure*}

An RFQ is a multi-purpose linear particle accelerator element able to
bunch (compress along the direction of movement) and accelerate a 
high-current ion beam, while keeping the beam tightly focused in the
transverse direction~\cite{kapchinskiiLINEARIONACCELERATOR1970, crandall:rfq, wanglerRFLinearAccelerators2008}. In an RFQ, an oscillating electric field is
generated between four \emph{vanes} (or \emph{rods}). 
The arrangement for one cell can be seen in Fig. ~\ref{fig:rfq_cell},
where the length is 
$\beta\cdot\lambda/2$, with $\beta$ the ratio of the beam velocity to the speed of light, and $\lambda$
the wavelength corresponding to the  electromagnetic wave driving the RFQ.
Typical RFQs comprise tens to hundreds of such cells,
each defined by three main parameters (the focusing strength $B$, 
the phase $\Phi$, and the modulation $m$). The IsoDAR RFQ is discussed in detail 
elsewhere~\cite{holtermann_technical_2021, koser_thermal_2021}. To optimize an RFQ,
the parameters of each cell have to be fine-tuned to 
yield desirable beam output qualities like a high beam transmission. In this paper, we call the RFQ parameters
\emph{design variables} (DVARs) and the beam output 
parameters \emph{objectives} (OBJs).

\subsection{The Julia Language}

Julia is an open-source dynamically typed programming language with  significant performance improvements over other languages common in scientific computing like Python and Matlab \cite{Julia-2017}. Julia's computational efficiency is especially apparent when performing costly calculations like neural network training, motivating our use of the language throughout this analysis and allowing for straightforward implementation of multi-threading and distributed computing to facilitate the completion of expensive computational tasks. In addition, Julia has built-in support for GPU programming using \texttt{CUDA.jl} \cite{besard2018juliagpu, besard2019prototyping}, making the language an obvious choice for building and training series of NNs on both local (CPU) and remote (GPU) machines.

%% file: Sec2_Methods.tex
\begin{figure*}[t!]
\begin{center}
\includegraphics[width=1.0\textwidth]
                {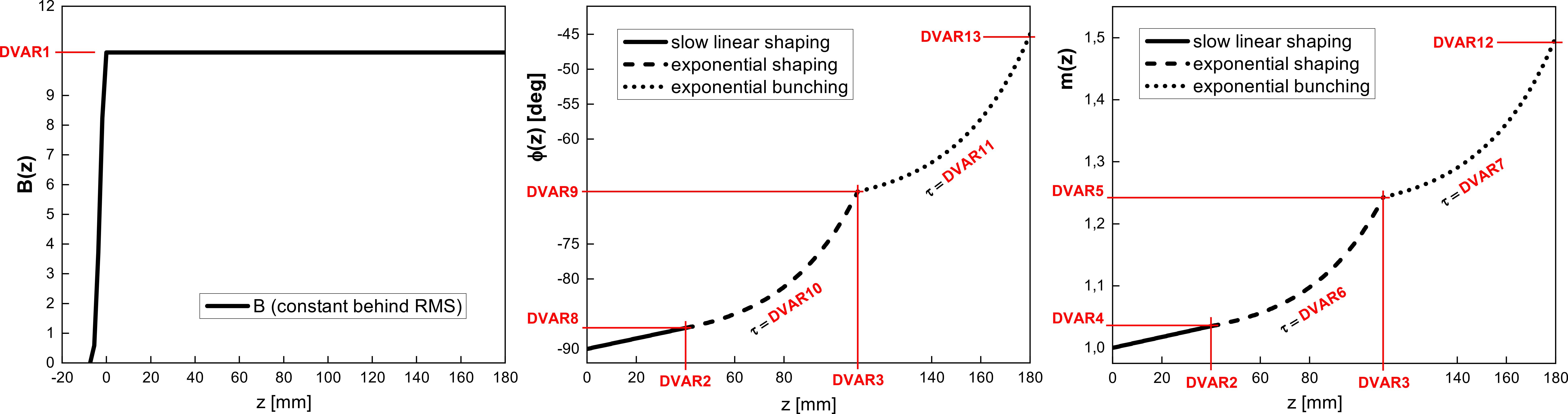}
\caption{The three main cell parameters (focusing function $B$,
         phase $\Phi$, and modulation $m$), parametrized along the length of the RFQ by fourteen DVARs (DVAR14, not shown, is the design energy determining the overall length). From 
         Ref.~\cite{frontiers}.}
\label{fig:cell_param}
\end{center}
\end{figure*}

\section{Methodology}
\label{sec:methods}

\subsection{Data Generation}
\label{sec:datagen}

In this study, we reuse the dataset from Ref.~\cite{frontiers}, consisting of 217,292 samples, each representing a randomly configured RFQ with corresponding output beam parameters obtained from beam dynamics simulations using the well-established code PARMTEQM~\cite{parmteqm}. The beam dynamics characteristics of an RFQ are fully described by a set of three main cell parameters ($B_n$, $\Phi_n$ and $m_n$) for each RFQ cell $n$, resulting in a total of $3n$ design variables. While historically the LANL Four-Section Procedure (FSP) has been used as a design strategy since the very first proof-of-principle RFQs at LANL at the end of the 1970s~\cite{crandall:rfq}, over the years, modified approaches have evolved to improve RFQ performance for specific applications. In the FSP, the RFQ is divided into four sections with dedicated functions: a radial matcher, a shaper, a gentle buncher and an accelerator, with the overall aim of smoothly matching the beam to the structure, bunching it, and accelerating it with as little 
losses as possible.

As a first approach for the IsoDAR RFQ, a baseline design was created via an adaptation of the FPS framework, but with special characteristics: (1)~in addition to the typical linear increase of the synchronous phase $\Phi$ and the modulation $m$ in the shaper (called the ``linear shaper"), a section of exponential increase is introduced (called the ``exponential shaper"); (2)~in the gentle buncher, not only the modulation but also the synchronous phase is ramped up exponentially (called the ``exponential buncher"); and (3)~the accelerator section is omitted due to the IsoDAR RFQ being intended for use as a dedicated pre-buncher.

For a machine learning based design optimization, a reduction of the number of the initially $3n$ design variables was pursued and thus a parametrization according to Fig.~\ref{fig:cell_param} was performed. This allowed to capture all crucial characteristics of the design functions while reducing the number of design variables to 14. While \texttt{DVAR1} corresponds to the value of the (constant) focusing strength and \texttt{DVAR2} and \texttt{DVAR3} set the lengths of the linear and exponential shaper, the total slope and smoothness of the shaping/bunching effect are characterized by \texttt{DVAR4}--\texttt{DVAR13}. \texttt{DVAR14} specifies the design energy at the RFQ exit, which determines the length of the RFQ.

Besides the 14 design variables (DVARs), the dataset from Ref.~\cite{frontiers} contains 6 objectives (OBJs): beam
transmission, output energy, RFQ length, and three beam emittances 
(one longitudinal, two transverse to the beamline). This data is summarized in Tab.~\ref{tab:datasummary}.

\begin{table}[t!]
\centering
\resizebox{\textwidth}{!}{%
 \begin{tabular}{|c l c c l|} 
 \hline
  Label & Variable & Lower Bound\textsuperscript{*} & Upper Bound & Physical Meaning \\
 \hline
 \texttt{DVAR1} & \texttt{Bmax} [1] & $8.5$ & $12.0$ & (Constant) focusing strength \\
 \texttt{DVAR2} & \texttt{mX1} [cm] & $5$ & $140$ & End position of linear shaper \\
 \texttt{DVAR3} & \texttt{mX2} [cm] & $\texttt{DVAR2} + 10$ & $160$ & End position of exponential shaper\\
 \texttt{DVAR4} & \texttt{mY1} [1] & $1.005$ & $1.7$ & Modulation at end of linear shaper\\
 \texttt{DVAR5} & \texttt{mY2} [1] & $\mathtt{DVAR4} + 0.05$ & $1.85$ & Modulation at end of exponential shaper\\
 \texttt{DVAR6} & \texttt{mtau1} [cm] & $1$ & $500$ & Exp.~param.~for modulation in exp.~shaper\\
 \texttt{DVAR7} & \texttt{mtau2} [cm] & $1$ & $500$ & Exp.~param.~for modulation in exp.~buncher\\
 \texttt{DVAR8} & \texttt{PhiY1} [deg] & $-89.95$ & $-30$ & Synchronous phase at end of lin.~shaper\\
 \texttt{DVAR9} & \texttt{PhiY2} [deg] & $\mathtt{DVAR8} + 2.5$ & $-25$ & Synchronous phase at end of exp.~shaper\\
 \texttt{DVAR10} & \texttt{Phitau1} [cm] & $1$ & $500$ & Exp.~param.~for sync. phase in exp.~shaper\\
 \texttt{DVAR11} & \texttt{Phitau2} [cm] & $1$ & $500$ & Exp.~param.~for sync. phase in exp.~buncher\\
 \texttt{DVAR12} & \texttt{mY3ref} [1] & $\mathtt{DVAR5} + 0.05$ & $2.0$ & Modulation at end of exp.~buncher\\
 \texttt{DVAR13} & \texttt{PhiY3ref} [deg] & $\mathtt{DVAR9} + 2.5$ & $-20$ & Synchronous phase at end of exp.~buncher\\
 \texttt{DVAR14} & \texttt{Eref} [MeV] & $0.055$ & $0.075$ & Design energy (at exit)\\
 \hline
 \texttt{OBJ1} & \multicolumn{3}{l}{Transmission [\%]} & \\
 \texttt{OBJ2} & \multicolumn{3}{l}{Output energy [MeV] ($E_{\textrm{out}}$)} & \\
 \texttt{OBJ3} & \multicolumn{3}{l}{RFQ length [cm]} & \\
 \texttt{OBJ4} & \multicolumn{3}{l}{Longitudinal emittance [MeV deg] ($\epsilon_{\textrm{long}}$)} & \\
 \texttt{OBJ5} & \multicolumn{3}{l}{$x$-emittance [cm mrad] ($\epsilon_x$)}  & \\
 \texttt{OBJ6} & \multicolumn{3}{l}{$y$-emittance [cm mrad] ($\epsilon_y$)} & \\
 \hline
 \end{tabular}}
 \caption{\label{tab:datasummary} Summary of design variables (DVARs) and objectives (OBJs) used for surrogate model training. \textsuperscript{*}In uniformly drawing random RFQ design variables, certain relationships between these parameters must be satisfied to ensure the RFQ is physical. This is further discussed in Sec.~\ref{sec:preprocessing}.}
\end{table}

\subsection{Data Preprocessing}
\label{sec:preprocessing}
Unlike Ref.~\cite{frontiers}, we perform transformations on the data set, as some variables have values that directly affect values of others. In particular, for some upper bounds $u_i$ and buffers $\delta_i$, $i \in \{3, 5, 9, 12, 13\}$, after a subset of the 14 DVARs for the $j$th sample are drawn from their respective uniform distributions, the rest are acquired via:
\begin{eqnarray} \label{eqn:correlatedDvars}
\mathtt{DVAR3}_j &\sim \textrm{Uniform}(\mathtt{DVAR2}_j + \delta_3, u_3) \nonumber\\
\mathtt{DVAR5}_j &\sim \textrm{Uniform}(\mathtt{DVAR4}_j + \delta_5, u_5) \nonumber\\
\mathtt{DVAR9}_j &\sim \textrm{Uniform}(\mathtt{DVAR8}_j + \delta_9, u_9) \\
\mathtt{DVAR12}_j &\sim \textrm{Uniform}(\mathtt{DVAR5}_j + \delta_{12}, u_{12}) \nonumber\\
\mathtt{DVAR13}_j &\sim \textrm{Uniform}(\mathtt{DVAR9}_j + \delta_{13}, u_{13}) \nonumber
\end{eqnarray}

\noindent With:
\begin{itemize}
\item $\delta_3$ the minimum allowable length of the exponential shaper,
\item $\delta_5$ the minimum allowable modulation increase in the exponential shaper,
\item $\delta_9$ the minimum allowable phase increase in the exponential shaper,
\item $\delta_{12}$ the minimum allowable modulation increase in the exponential buncher, and
\item $\delta_{13}$ the minimum allowable phase increase in the exponential buncher.
\end{itemize}

\noindent Values of the $\delta_i$ and $u_i$ can be inferred from Tab.~\ref{tab:datasummary}. Having these buffers ensures that physically realistic RFQ sections are created. For instance, if $\mathtt{DVAR3}_j \sim \textrm{Uniform} (\mathtt{DVAR2}_j, u_3)$, it is possible that a randomly generated RFQ would have an exponential shaper component of zero length, which is nonphysical for the family of accelerators considered in this study.

As a result, design variables like $\texttt{DVAR3}$ are not uniformly distributed. We can introduce a variable $\texttt{DVAR3}'$ that is uniformly distributed according to the following transformation of $\texttt{DVAR2}, \texttt{DVAR3}$, and constants:

\begin{equation}
\label{eqn:dvar3transform}
\mathtt{DVAR3}'_j = \frac{\mathtt{DVAR3}_j - (\mathtt{DVAR2}_j + \delta_3)}{u_3 - (\mathtt{DVAR2}_j + \delta_3)}
\end{equation}

\noindent And likewise for the remaining pairs of features. A side-by-side comparison of distributions of \texttt{DVAR3} and the transformed \texttt{DVAR3}' is shown in Fig.~\ref{fig:dvar3transform} in \ref{sec:transformedhist}.

\begin{figure*}[t!]
    \centering
    \includegraphics[width=0.75\textwidth]{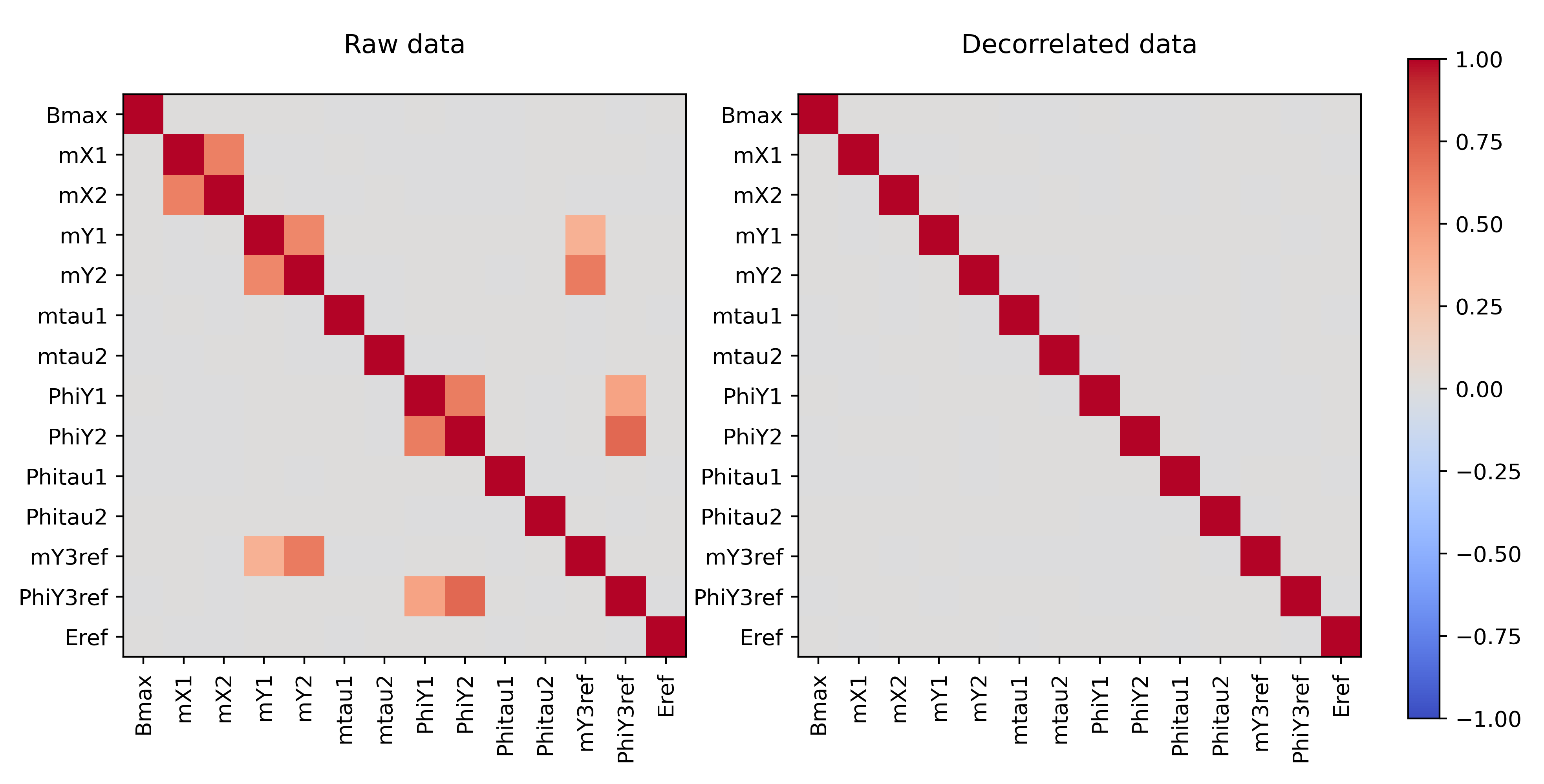}
    \caption{Correlation matrices for raw and decorrelated feature data (DVARs).}
    \label{fig:corrmtx}
\end{figure*}

Correlation matrices for the 14 design variables before and after the described pre-processing transformation of necessary DVARs are shown in Fig.~\ref{fig:corrmtx}. We call these transformed design variables decorrelated, since all correlations have been removed. Each feature (after the decorrelating step) was later scaled to have minimum $-1$ and maximum $+1$.

To assess the impact of decorrelating data on NN training and performance, we trained two identical neural networks on both correlated and decorrelated scaled datasets. We observed no significant difference in the training times or prediction accuracy between these two NNs; this is to be expected, NNs are, in general, over-parameterized so decorrelation of features is not generally a necessary step. Eventually, however, the top-performing surrogate model in this study will serve as an important component of a design optimization. It is easier in a prebuilt Bayesian optimization package to enforce the relationships between these DVARs listed in Eqs.~\ref{eqn:correlatedDvars} by working with decorrelated features; we can perform a multi-objective design optimization under no additional constraints (besides the constraint that the scaled, decorrealted features should have minima $-1$ and maxima $1$). All NNs we present in the remainder of this study are trained on the uncorrelated data, and we recommend decorrelating the design variables as a best practice moving forward, particularly when built surrogate models will be plugged into any kind of optimization scheme.

\subsection{Number of RFQ Cells as a Previously Hidden Variable}
\label{sec:rfqcellnumber}

The models reported in Ref.~\cite{frontiers} and early NNs built in this study have one thing in common: predictions of transverse emittances (\texttt{OBJ5} and \texttt{OBJ6}) were significantly less accurate than those of the other objective variables. In Refs.~\cite{guptaImprovingSurrogateModel2021, winklehner:njp} the
authors also observe this for a cyclotron and the LCLS-II injector beam line, making the failure to accurately predict RMS emittance a problem worth exploring. In \ref{sec:emittancejointdists}, we show that for surrogate models similar to those presented in Ref.~\cite{frontiers}, patterns in the true and surrogate-model-predicted emittances indicate some underlying hidden variable should be included to resolve the asymmetry between the two transverse directions.

\begin{figure*}[t!]
    \centering
    \includegraphics[width=0.6\textwidth]{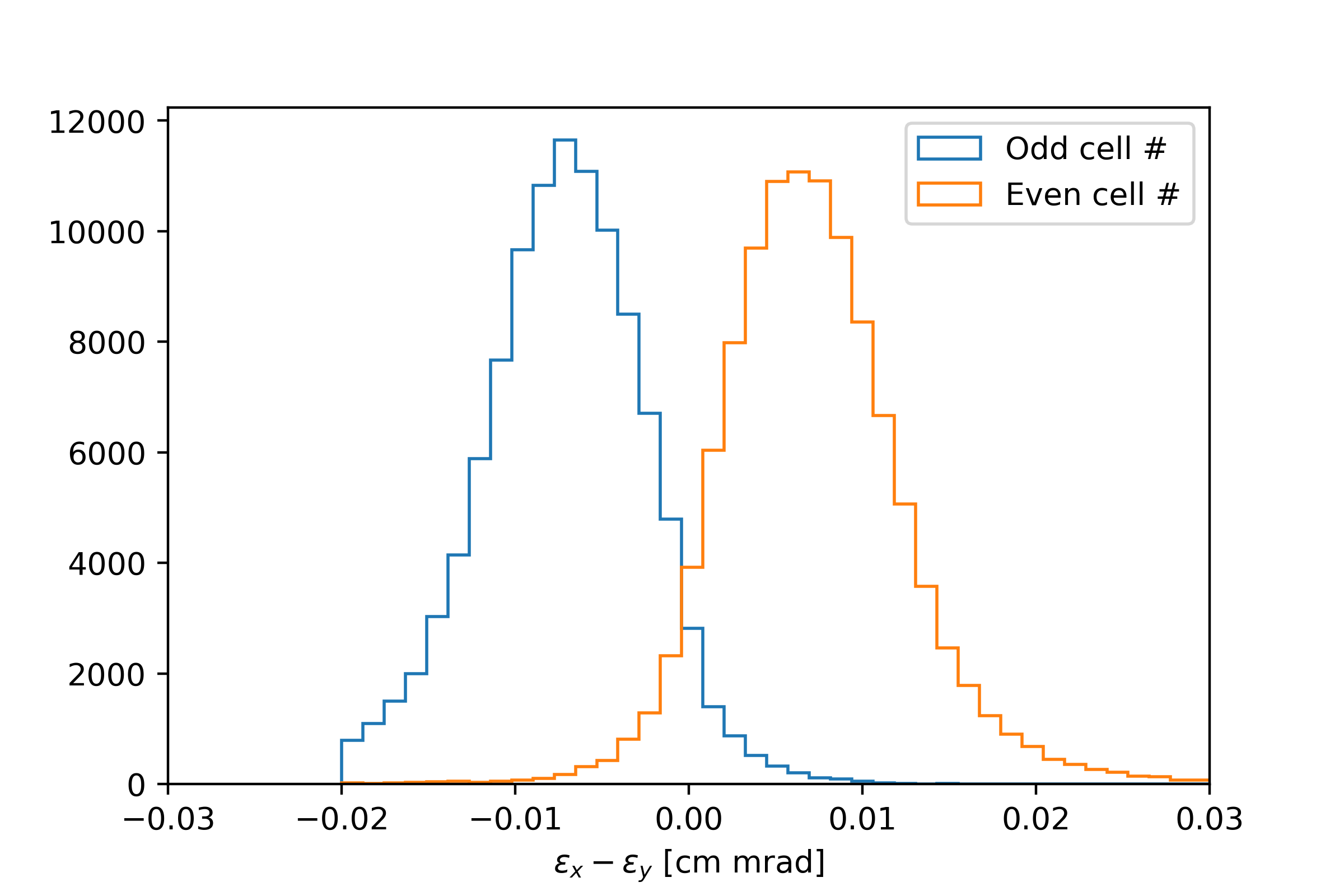}
    \caption{Overlaid distributions of the difference in transverse emittances for samples having odd and even cell numbers. The clear separation of these distributions indicates the inclusion of the number of cells is vital to the prediction of transverse emittance.}
    \label{fig:oddevenpredictemit}
\end{figure*}

To this end, we extend the the dataset from Ref.~\cite{frontiers} to include a fifteenth feature used for training the surrogate models produced in this work; namely, whether the number of RFQ cells
is even or odd. As discussed in Sec.~\ref{sec:rfq}, across any particular RFQ cell, the beam is simultaneously focused in one of the two transverse directions while being defocused in the other. The number of RFQ cells (specifically whether the number of RFQ cells is even or odd, which we denote as the ``parity" of the RFQ in this work) is essential in resolving any asymmetries in the beam characteristics across the two transverse directions, such as the two transverse emittances $\epsilon_x$ and $\epsilon_y$. Histograms of $\epsilon_x - \epsilon_y$ for simulated RFQs with odd and even numbers of cells are shown in Fig.~\ref{fig:oddevenpredictemit}, further elucidating this concept. We added an extra column to the features: a binary variable denoting whether the total number of RFQ cells is odd or even.

\begin{figure*}[t!]
    \centering
    \includegraphics[width=0.6\textwidth]{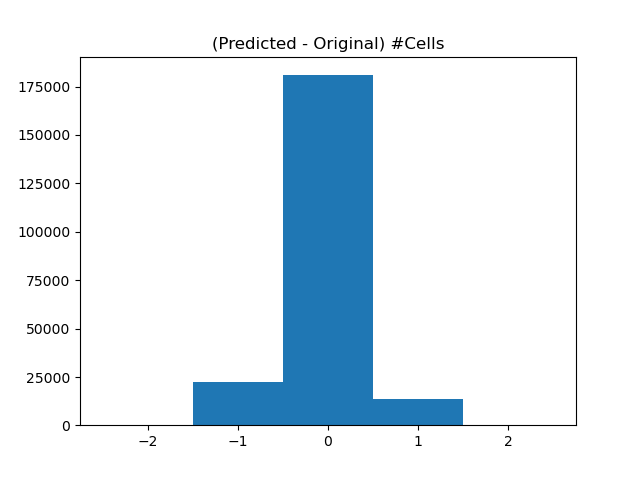}
    \caption{Comparison of algorithmic predictions for the total number of RFQ cells based on the 14 other DVARs to ground truth values.}
    \label{fig:cellnumberalgoresults}
\end{figure*}

This ``15th DVAR'' can be obtained from the other 14 DVARs
in a fully deterministic way, and this calculation becomes another preprocessing step of our data. The algorithm that we devised 
in Python is based on the work of Kapchinskii and Teplyakov~\cite{kapchinskiiLINEARIONACCELERATOR1970} as well as 
Crandall and Wangler~\cite{crandallEndingRFQVanetips1994, crandallPARMTEQBeamDynamics1988, crandallRFQRadialMatching1984, crandallRFQuadrupoleBeam1979, wanglerRFLinearAccelerators2008}, 
 calculating an ideal particle's energy gain for each RFQ cell, based on the parametrization
of cell parameters shown in Fig.~\ref{fig:cell_param}. Each preceding
cell thus determines the position along the $z$-axis at which to evaluate the parametrization curves for the next cell.
We tested the algorithm on all data in the training and validation
sets and obtained a $>83\%$ success rate in predicting even or odd
number of cells (cf. Fig.~\ref{fig:cellnumberalgoresults}).
\footnote{Feeding NNs the raw number of cells, rather than the binary parity variable, experimentally proved difficult for the NNs to use when determining the sign of $\epsilon_x - \epsilon_y$. We suspect that this is due to the fact that the number of cells in the virtual RFQs in this study is $\sim \mathcal{O}(100)$, so performing the uniform-minimum-and-maximum transformation degrades the ability of the network to resolve whether the number of cells was originally odd or even.} The code for this routine is available in the \texttt{compute\_cellnumber/compute\_cellnumber.py} file on our GitHub~\cite{rfqnn-github}. The $\approx17$\% miss rate can be attributed to the fact that
our Python script uses approximations for each full cell, while
Pari, an RFQ design code used in sequence with PARMTEQM, integrates over the cell numerically by slicing it.
A much slower, but equally legitimate way of calculating DVAR15 
during preprocessing would be to perform a system call to 
\texttt{Pari.exe} and to read the corresponding output file. This would
 predict the parity of the cell number with $100\%$ 
success. This is how the training set was generated. For the
optimizer, we chose the fast Python-based algorithm as it is
not restricted to Windows platforms and not export-controlled in the USA
(i.e. only allowed to be used without restrictions within the USA).

It is important to note that the inclusion of this additional design variable does not impact the integrity of the surrogate model, nor does it significantly increase our surrogate model's computational overhead. The algorithm 
is fast-executing and the 15th DVAR follows from the other 14
DVARs fully deterministically. All NNs are trained on correct parity information; the approximation algorithm is only used during the optimization phase. During the optimization process, the 
use of the Python algorithm, which is incorrect 
$\approx17\%$ of the time, may lead to slightly longer time-to-solution and a larger error associated with the emittance
results, however, this error is comparable with the NN MAPEs for the emittance in general. Furthermore, 
each point in the Pareto-front is ultimately calculated by a call to the Surrogate Model. We observed empirically that artificially swapping the parity from $+1$ to $-1$ has the effect of (approximately) swapping the predictions of $\epsilon_x$ and $\epsilon_y$
(i.e. moving from one branch to the other in Fig.~\ref{fig:trueemits}), while keeping predictions on the other 4 objectives constant. Since the distribution
of even and odd in the incorrectly calculated parities is 
about equal, the Pareto front will look approximately the same. Finally, the chosen Pareto-optimal point after optimization will have to be corroborated by the original
simulation software, where incorrect parity predictions will be evident. 
The designer can modify the machine accordingly. 
Even with a $17\%$ miss rate in predicting the parity of an RFQ, we still achieve substantial performance improvements over previous efforts to predict emittances.

\subsection{Surrogate Model Reporting}
\label{sec:surrogatemodelreporting}
Neural network performance is evaluated by computing $R^2$ scores and mean absolute percent errors (MAPEs). We cite the ``aggregate $R^2$" score,  the average of the $R^2$ scores for each of the six response variables, to report the total accuracy of the model. This does not, however, give any insight into predictive accuracies of each of the six objectives separately: we use MAPEs on the unscaled data to handle this task.

\subsection{Training on GPUs}
\label{sec:gpu}
As part of our efforts to fully explore the Julia language, and to exploit the significant speedup coming from massive parallelization of matrix algebra, we migrated NN training to GPUs. 
Here we used two NVIDIA A30 Tensor Core GPU cards with 24~GB of VRAM 
each, combined with Intel Xeon(R) Silver 4310 12-core CPUs (we assigned one thread per GPU) on the MIT ``SubMIT'' 
cluster~\cite{SubMITGettingPhysics2022} for training ``RFQNet1'' and 
``RFQNet2'' (see below). We used CUDA version 11.6.
While we do not present a GPU benchmark, we can point out that
the speedup over CPU training was significant.


\begin{table}[t!]
\centering
 \begin{tabular}{|c c|} 
 \hline
 Hyperparameter & Value(s) scanned \\
 \hline
 NN depth & 4, 5, 6 \\ 
 NN width & 50, 75, 100 \\
 Batch size & 1024 \\
 Learning rate & 0.001 \\
 Activation function & Sigmoid \\
 Loss function & Mean squared error \\
 \hline
 \end{tabular}
 \caption{\label{tab:hyperparamscanvalues}Values chosen for initial hyperparameter scan in Julia.}
\end{table}

\subsection{Implementation of Neural Network Training and Hyperparameter Scans}
\label{sec:hyperparameterscan}
Operating on the full decorrelated dataset, we use Julia to implement grid search scans over NN architecture hyperparameters; namely, width and depth. We hold batch size, learning rate, and activation functions constant for this study but plan to expand future searches to these hyperparameters, as well. All NNs were trained using mean squared error as the loss function. Scanned NN hyperparameters are summarized in Tab.~\ref{tab:hyperparamscanvalues}. Neural network training was performed primarily using the \texttt{Flux.jl} package, an open-source Julia library used for training deep ML models Ref.~\cite{Flux.jl-2018, innes:2018}. NNs of different architectures were trained in parallel to reduce script runtime. Results from Ref.~\cite{frontiers} indicate that neural networks trained with the largest batch sizes were desirable. While larger batch sizes may be preferred in principle to show NNs more data per training step, they ultimately increase total training time \cite{batchsizes}. The use of Julia allows us to feasibly consider training neural networks with larger batch sizes in a more reasonable amount of time. Hence, we choose to hold batch size constant at 1024, outside of the range of that explored by Ref.~\cite{frontiers}. While it is difficult to interpret why the surrogate models developed in this study seemed to prefer larger batch sizes, we suspect that because we are sampling the $14$-dimensional feature space uniformly, each training step sees a more representative sample of the training data, allowing our networks to generalize better. Also, to fully leverage the parallelization capabilities of GPUs used for training these NNs, larger batch sizes are computationally preferred.

Data was split into training and test sets (of proportions 80\% and 20\%, respectively). The test set was withheld from any analysis for the entirety of the scan, and was only used for computing the final test-set MAPEs shown in Tab.~\ref{tab:valr2}. We used the \texttt{MLUtils.jl} package (Ref.~\cite{mlutilsgithub}) to perform 5-fold cross-validation, wherein each neural network was trained on an 80\% subsample of the training data, with the remaining 20\% used as a validation dataset to estimate out-of-sample model performance. NNs were not trained on this smaller subset of data. The validation sets used in each fold of the cross-validation procedure were disjoint.

Each neural network was constructed using a sigmoid activation function. The \texttt{Flux.jl} ADAM (Ref.~\cite{Kingma2015AdamAM}) optimizer was used to perform stochastic gradient descent in neural network training, using the default learning rate of $0.1\%$ unless otherwise specified. All neural networks were trained for 2500 epochs, which appeared experimentally to be a reasonable threshold for convergence (one in which improvements in predictive accuracy began to plateau while preserving agreement between the validation and training set prediction errors). The results of this hyperparameter scan are reported in 
Sec.~\ref{sec:results1}, and the best performing of these networks is referred to as ``RFQNet1". This network is summarized in table Tab.~\ref{tab:modelsummaries}.

In prototyping of the surrogate models developed in this study, we also added dropout regularization to trained neural networks and tested different combinations of dropout rates and learning rates. The inclusion of dropout regularization, in general, negatively impacted the validation-set predictive accuracy of the networks in the worst case, while different objectives preferred different combinations of dropout and learning rates, in the best case. None of the tested networks outperformed RFQNet1 and RFQNet2 presented in this work.

\subsection{Additional Design-Motivated Data Preprocessing}
\label{subsec:transmissioncut}
The RFQ dataset used both in this study as well as in Ref.~\cite{frontiers} was generated by performing basic random sampling of the 14 design parameters within their respective specified value ranges, and determining the objective variables by simulating beam traversing the RFQ using a physics engine. However, generating RFQs in this way can result in RFQs that are especially undesirable and even unrealistic. Many of the virtual RFQs were determined to have an alarmingly low beam transmission of $60\%$ or less. In the case that an accelerator engineer would like to use a surrogate model to optimize their RFQ design, it is reasonable to assume their RFQ's beam transmission will be much higher. We therefore perform the same hyperparameter scan outlined in Sec.~\ref{sec:hyperparameterscan} on data culled to have simulated beam transmissions of at least $60\%$. The best-performing neural network of this hyperparameter scan is referred to as ``RFQNet2". RFQNet2's structure is compared to that of RFQNet1 in Tab.~\ref{tab:modelsummaries}.

Due to RFQs being complex machines with many cells, and parameters like transmission 
depending on the interplay between the DVARs rather than being smooth monotonically changing functions, it is not possible to determine ab-initio which DVARs need to be restricted (and where, as there are many local minima) to limit transmission to $>60\%$. Thus, we cut based on the OBJ rather than the DVARs. Chosing a value of $60\%$, which is
still far away from the desired transmissions of RFQs, gives the surrogate model
enough margin for us to be able to perform optimizations.

\subsection{RFQ Design Optimization}
\label{sec:designoptimization}
To demonstrate the utility of such ML-powered surrogate models, we used RFQNet2 as the surrogate model for the beam dynamics simulator to run an optimization of the design parameters of an RFQ like the one proposed in the IsoDAR experiment. Such an RFQ has the following optimal properties:

\begin{itemize}
\setlength\itemsep{0.01em}
\item Maximal beam transmission
\item Minimal difference between output and target (70 KeV) energies
\item Minimal RFQ length
\item Minimal longitudinal emittance
\item Minimal transverse emittance
\end{itemize}

Finding solutions to such a multidimensional optimization problem motivates the use of Bayesian optimization. The goal of such a procedure is to find the set of RFQs that are \textit{Pareto-optimal}, where all points in the Pareto front are optimal in the sense that improvements in some objectives come at the cost of others. To this end, we make use of the package \texttt{Surrogates.jl}, as part of the open-source Scientific Machine Learning Initiative of Julia \cite{rackauckas2017differentialequations, rackauckas2020universal}. This package allows us to use our surrogate model as part of the acquisition function in the optimization algorithm to select and evaluate data points in the feature space. As the algorithm selects each 14-dimensional data point to test for optimality, we take the intermediate step of employing the procedure outlined in Sec.~\ref{sec:rfqcellnumber} to determine whether the number of RFQ cells will be odd or even to extend the feature vector with a fifteenth entry.
To narrow the Pareto-optimal solution space to a set of RFQ design parameters feasible for the goals of IsoDAR's RFQ, we search for points in the Pareto set with beam transmission of at least $90\%$ and as low as possible transverse beam emittance ($\epsilon_x, \epsilon_y \leq 0.04$~cm mrad). Since longitudinal and transverse beam emittances are correlated, we expect that the longitudinal beam emittance will similarly be low.

%% file: Sec3_Results.tex
\section{Results}
\label{sec:results} 

\begin{figure*}[t!]
    \begin{center}
        \includegraphics[width=0.95\textwidth]{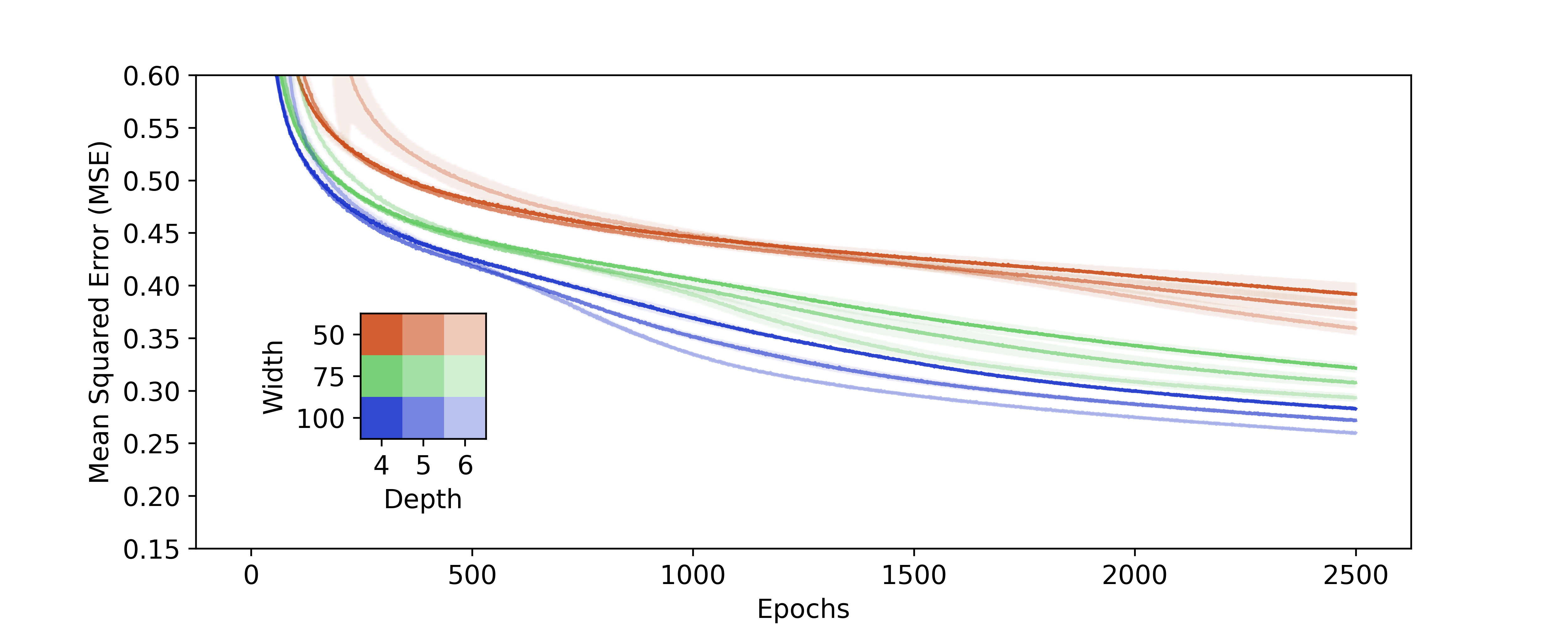}
    \end{center}
    \begin{center}
        \includegraphics[width=0.95\textwidth]{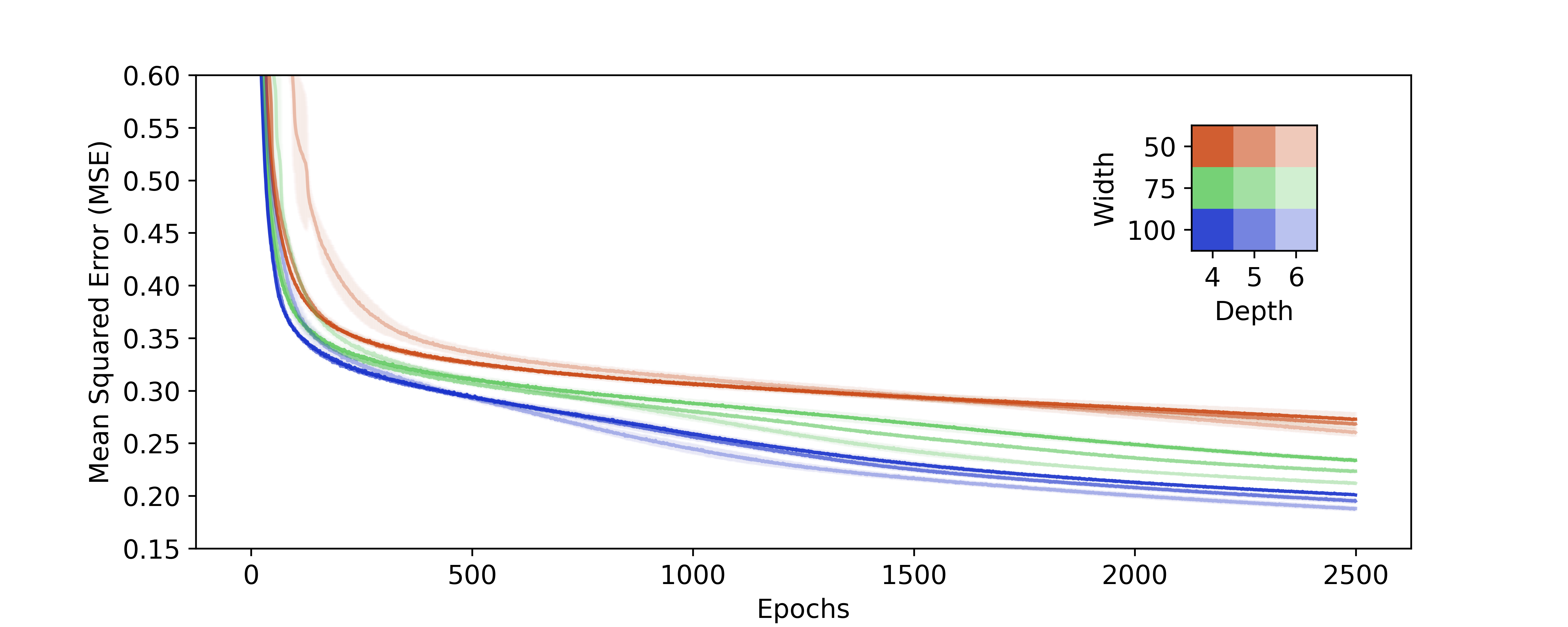}
        \caption{Training set loss curves (mean squared error) for scanned neural network architectures (Tab.~\ref{tab:hyperparamscanvalues}), trained on the complete dataset (top) and samples having beam transmission $\geq 60\%$ (bottom). The solid line in the center of each curve represents the cross-fold loss mean, and one standard deviation is shaded above and below.}
        \label{fig:losses}
    \end{center}
\end{figure*}

\begin{table*}[htp!]
\centering
 \begin{tabular}{|c |c c c|} 
 \hline
 & Depth 4 & Depth 5 & Depth 6 \\
 \hline
 Width 50 & $0.9884\pm 0.000311$ & $0.9887\pm 0.000217$ & $0.9893\pm 0.000127$ \\
 Width 75 & $ 0.9900\pm 0.000169$ & $0.9904\pm 0.0000864$ & $0.9907\pm 0.000145$ \\
 Width 100 & $0.9905\pm 0.0000763$ & $0.9908\pm 0.0000698$ & $0.9908\pm 0.00017$ \\
 \hline
 \multicolumn{4}{c}{} \\
 \hline
 & Depth 4 & Depth 5 & Depth 6 \\
 \hline
 Width 50 & $0.9901\pm 0.0002$ & $0.9902\pm 0.000174$ & $0.9905\pm 0.00026$ \\
 Width 75 & $0.9912\pm 0.000129$ & $0.9915\pm 0.000178$ & $0.9918\pm 0.000217$ \\
 Width 100 & $0.9918\pm 0.0000719$ & $0.9917\pm 0.000152$ & $0.9920\pm 0.0000623$ \\
 \hline
 \end{tabular}
 \caption{\label{tab:valr2} Aggregated validation-set $R^2$ scores for each set of hyperparameters (Tab.~\ref{tab:hyperparamscanvalues}), for NNs trained on the complete dataset (top) and samples having transmission $\geq 60\%$ (bottom). Each $R^2$ score is computed across all objective variables, but is not representative of the prediction accuracy of individual objectives; MAPEs for each objective for each NN architecture are shown in \ref{sec:mapebyobjectiveappendix}. Small apparent differences in validation-set $R^2$ scores correspond to a few percentage-point differences in MAPEs.}
\end{table*}

\subsection{Results of the Two Preliminary Hyperparameter Scans}
\label{sec:results1}
Loss curves for the full-dataset training of the 9 different neural network architectures are shown in Fig.~\ref{fig:losses}, top, and the validation set aggregate $R^2$ scores are summarized in Tab.~\ref{tab:valr2}, top. The NN with the highest validation set aggregate $R^2$ is the neural network with depth 5 and width 100, which we select over the $6 \times 100$ network due to the lower cross-fold standard deviation. This $5 \times 100$ neural network is referred to as RFQNet1. Complete validation set MAPEs for all neural networks generated in this scan are summarized in Tab.~\ref{tab:notransmission-cut-mapes-by-obj} in \ref{sec:mapebyobjectiveappendix}.

As discussed in Sec.~\ref{subsec:transmissioncut}, limiting the dataset used in training and testing NNs to samples with a transmission of at least $60\%$ is physically well-motivated. Loss curves for each of the neural networks trained on the data subset having transmission $\geq60\%$ are shown in Fig.~\ref{fig:losses}, bottom, and the validation-set aggregate $R^2$ scores are summarized in Tab.~\ref{tab:valr2}, bottom. The depth 6 and width 100 neural network had the highest aggregate $R^2$ score, and is referred to as RFQNet2. The complete validation set MAPEs for similar NNs scanned are summarized in Tab.~\ref{tab:transmission-cut-mapes-by-obj} in \ref{sec:mapebyobjectiveappendix}.

\begin{figure*}[t!]
    \begin{center}
        \includegraphics[width=0.9\textwidth]{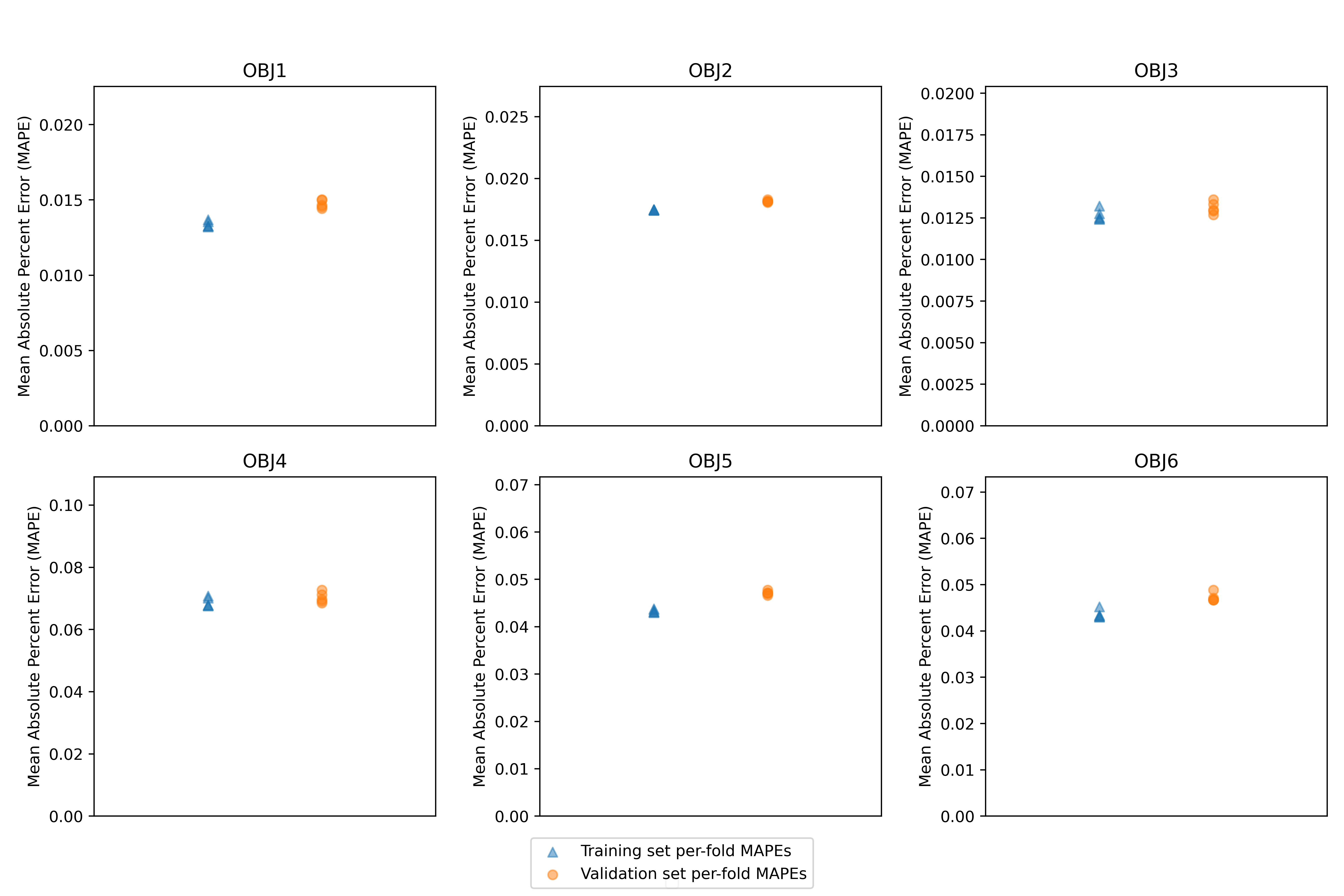}
    \end{center}
    \begin{center}
        \includegraphics[width=0.9\textwidth]{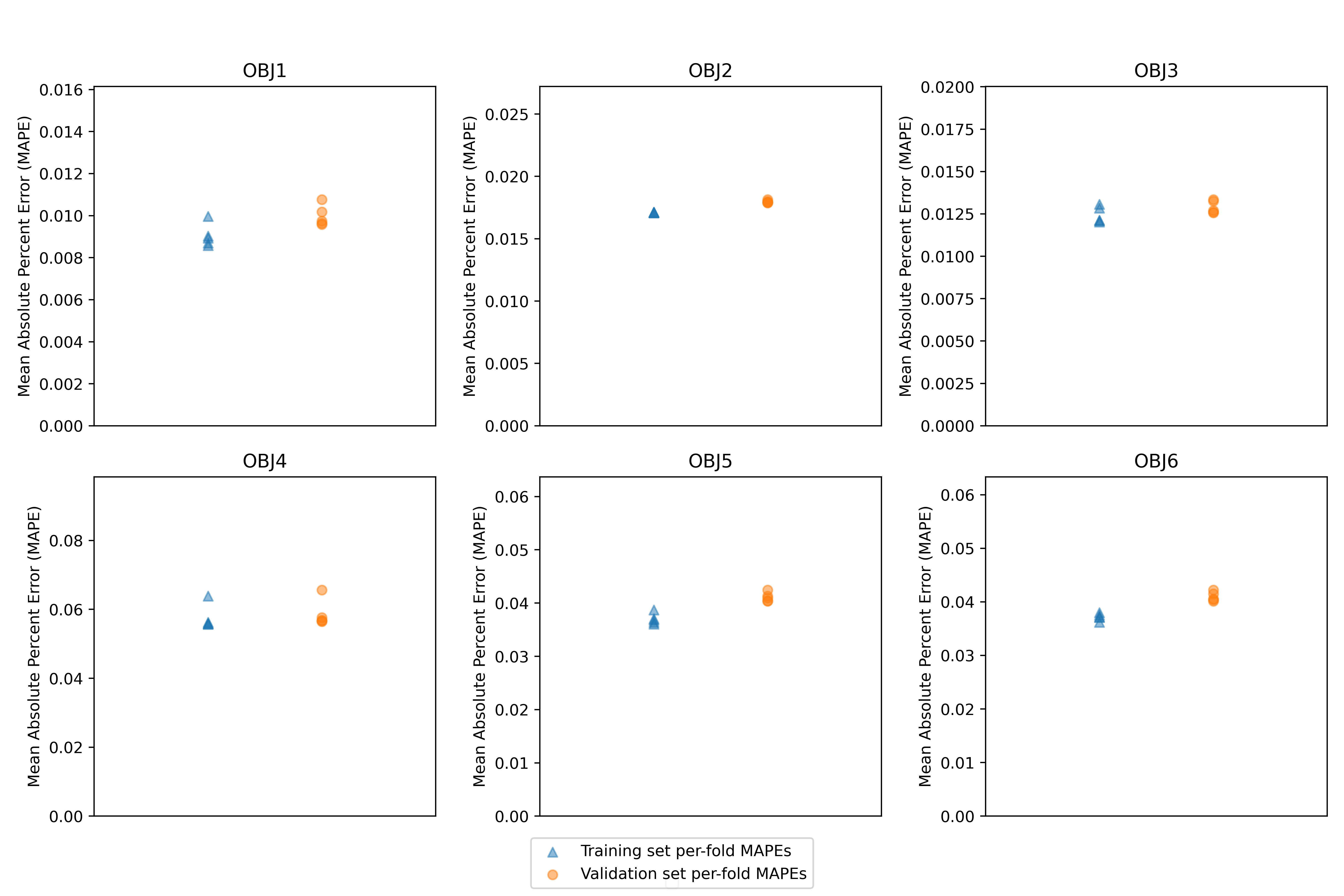}
        \caption{Comparison of mean absolute percent errors (MAPEs) on each of the 5 performed cross-validation folds on training and validation datasets for RFQNet1 (top) and RFQNet2 (bottom). RFQNet1/RFQNet2 are the optimal networks whose defining characteristics are summarized in Tab.~\ref{tab:modelsummaries}. Nearly equivalent MAPEs for both training and validation sets point to models that are unlikely to be overfit to the training data.}
        \label{fig:overfittingcheck}
    \end{center}
\end{figure*}

The test set MAPEs for both RFQNet1 and RFQNet2 are compared to the best-performing model from Ref.~\cite{frontiers} in Tab.~\ref{tab:mapes}. Checks for overfitting to the data are shown by reporting the training and validation set MAPEs for each of the 5 folds performed in the cross validation for choosing the optimal architectures of the NNs, Fig.~\ref{fig:overfittingcheck}. The MAPEs for the training and validation sets are generally close enough that we can conclude that neither model is overfit to the training data. Since RFQNet2's MAPEs for each of the 6 objectives are equal to or lower than those of RFQNet1, RFQNet2 should be chosen over RFQNet1 for use as a surrogate model in a design optimization.

\begin{table*}[t!]
\centering
\resizebox{\textwidth}{!}{%
\begin{tabular}{|c c c c l|}
\hline
\, & Depth & Width  & Validation-Set $R^2$ & Additional Specifications \\
\hline
RFQNet1 & $5$ & $100$ & $0.9908 \pm 0.0000698$ & Used complete training dataset\\
RFQNet2 & $6$ & $100$ & $0.9920 \pm  0.0000623$ & Subset of data having transmission $\geq 60\%$\\
\hline
\end{tabular}}
\caption{\label{tab:modelsummaries} Summary of RFQNet1 and RFQNet2, whose architectures where chosen by the cross-validation results of a hyperparameter scan over neural network depth and width. $R^2$ scores correspond to those reported in Tab.~\ref{tab:valr2}.}
\end{table*}

\begin{table*}[t!]
\centering
 \begin{tabular}{|clccc|} 
 \hline
Label & Objective Variable & RFQNet1 & RFQNet2  &
Ref.~\cite{frontiers} \\ 
 \hline
 \texttt{OBJ1} & Transmission [\%] & $1.5\%$ & $0.97\%$ & $2.4\%$ \\
 \texttt{OBJ2} & E$_\mathrm{out}$ [MeV] & $1.8\%$ & $1.8\%$ &$1.9\%$ \\
 \texttt{OBJ3} & RFQ length [cm] & $1.3\%$ & $1.3\%$ & $2.0\%$ \\
 \texttt{OBJ4} & $\epsilon_\mathrm{long.}$ [MeV-deg] & $6.9\%$ & $5.8\%$ & $8.2\%$ \\
 \texttt{OBJ5} & $\epsilon_x$ [cm-mrad] & $4.8\%$ & $4.1\%$  & $12.8\%$ \\
 \texttt{OBJ6} & $\epsilon_y$ [cm-mrad] & $4.8\%$ & $4.0\%$ & $12.5\%$ \\
 \hline
 \end{tabular}
 \caption{\label{tab:mapes} Test set MAPEs for each of the 6 objective variables studied, compared to previous results in Ref.~\cite{frontiers}. Specifications of RFQNet1 and RFQNet2 are given in Tab.~\ref{tab:modelsummaries}. These NNs were chosen via the hyperparameter scans summarized in Tab.~\ref{tab:valr2}, top and bottom, respectively.}
\end{table*}

\subsection{Design Optimization of IsoDAR's RFQ}
\begin{figure*}[htp!]
\centering
\includegraphics[width=1.0\textwidth]
{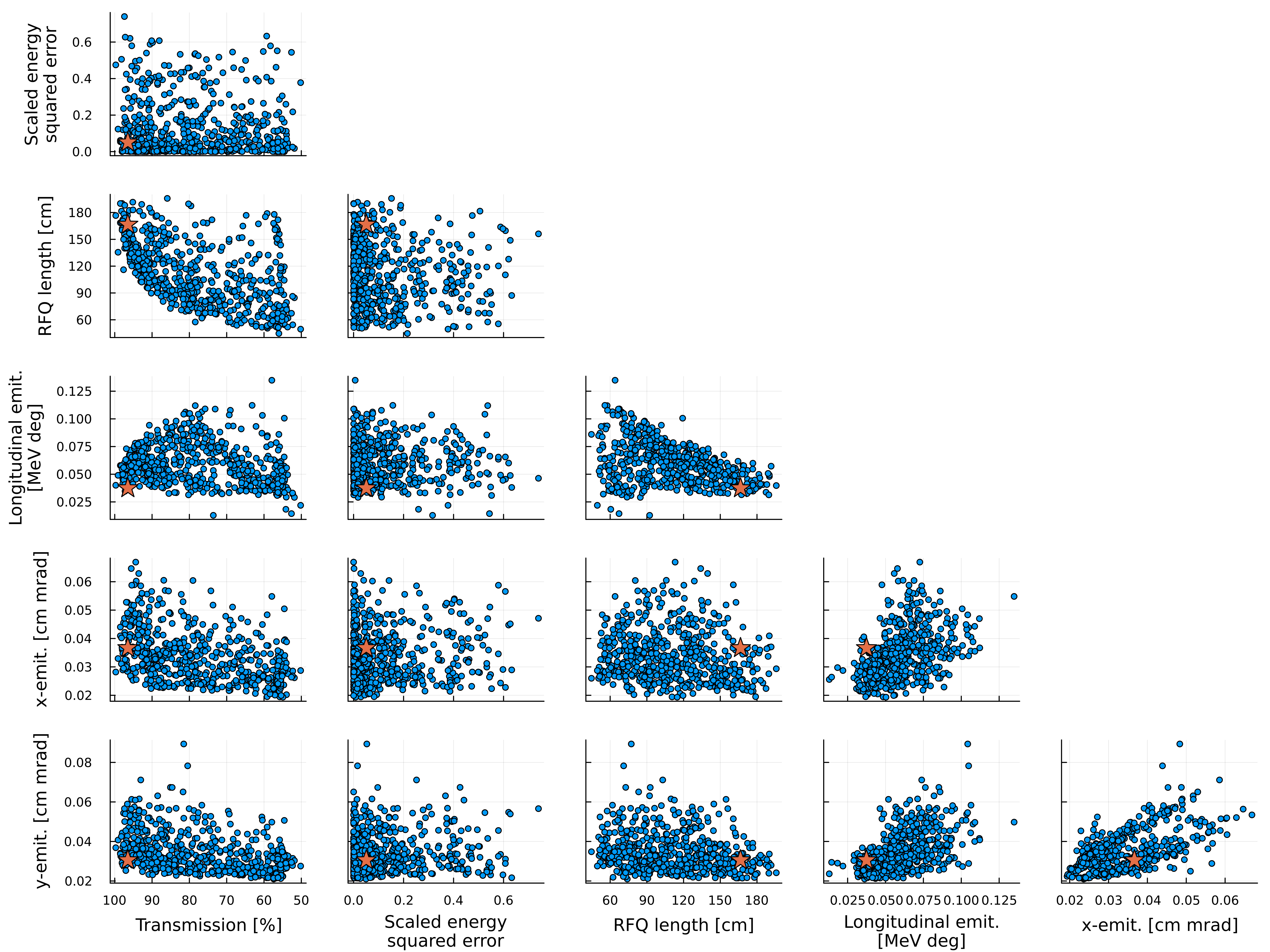}
\caption{The RFQ design optimal for IsoDAR's use case was computed by running a multi-objective Bayesian optimization scheme using the best-performing neural network as the surrogate model for the algorithm's acquisition function. 2D projections of the 6D Pareto front are shown above, with the optimal design configuration (see Tab.~\ref{tab:optimal-design}) marked with an orange star.}
\label{fig:pareto_plot}
\end{figure*}

As mentioned in Sec.~\ref{sec:designoptimization}, we use RFQNet2 as a surrogate model for the multi-objective Bayesian Optimization of an RFQ suitable for IsoDAR's needs. This model was a depth 6, width 100 neural network trained on data having to beam transmissions of $\geq 60\%$. 2D projections of the 6D Pareto front are plotted in Fig.~\ref{fig:pareto_plot}. Of the hundreds of points computed to be Pareto-optimal, we narrow down the possible designs to ones having transmissions of at least $95\%$ with radial and longitudinal beam emittances below $0.04 \textrm{ cm-mrad}$ and $0.04 \textrm{ MeV }^\circ$, respectively. We then weigh these remaining options according to their overall performance in energy matching and minimal RFQ length. The optimum's design variables and corresponding beam output parameters from this study are summarized in Tab.~\ref{tab:optimal-design}.

It is important to note that the boundaries and relationships between the first 14 design variables were chosen to ensure physical realizability of an RFQ. Because these were folded into the optimization procedure (via the decorrelation procedure described in Sec.~\ref{sec:preprocessing}), the optimal RFQ is indeed a physically realistic design.

\begin{table}[t!]
\centering
\resizebox{\textwidth}{!}{%
\begin{tabular}{|c l c c|}
\hline
Label & Design variable & \multicolumn{2}{c|}{Optimal value} \\
\hline
\texttt{DVAR1} & \texttt{Bmax} [1] & \multicolumn{2}{c|}{$9.1289$} \\
\texttt{DVAR2} & \texttt{mX1} [cm] & \multicolumn{2}{c|}{$46.1325$} \\
\texttt{DVAR3} & \texttt{mX2} [cm] & \multicolumn{2}{c|}{$92.6490$} \\
\texttt{DVAR4} & \texttt{mY1} [1] & \multicolumn{2}{c|}{$1.0430$} \\
\texttt{DVAR5} & \texttt{mY2} [1] & \multicolumn{2}{c|}{$1.4656$} \\
\texttt{DVAR6} & \texttt{mtau1} [cm] & \multicolumn{2}{c|}{$425.9281$} \\
\texttt{DVAR7} & \texttt{mtau2} [cm] & \multicolumn{2}{c|}{$160.8364$} \\
\texttt{DVAR8} & \texttt{PhiY1} [deg] & \multicolumn{2}{c|}{$-82.9247$} \\
\texttt{DVAR9} & \texttt{PhiY2} [deg] & \multicolumn{2}{c|}{$-68.7336$} \\
\texttt{DVAR10} & \texttt{Phitau1} [cm] & \multicolumn{2}{c|}{$199.8206$} \\
\texttt{DVAR11} & \texttt{Phitau2} [cm] & \multicolumn{2}{c|}{$176.4324$} \\
\texttt{DVAR12} & \texttt{mY3ref} [1] & \multicolumn{2}{c|}{$1.6405$} \\
\texttt{DVAR13} & \texttt{PhiY3ref} [deg] & \multicolumn{2}{c|}{$-56.4804$} \\
\texttt{DVAR14} & \texttt{Eref} [MeV] & \multicolumn{2}{c|}{$0.0723$} \\
\hline
\multicolumn{4}{c}{} \\
\hline
Label & Objective variable & Predicted value & Confirmed value (\textit{\% error}) \\
\hline 
\texttt{OBJ1} & Beam transmission [\%] & $96.4862$ & $97.14$ ($\mathit{0.67\%}$)\\
\texttt{OBJ2} & Output energy [MeV] & $0.0737$ & $0.07350$ ($\mathit{0.27\%}$) \\
\texttt{OBJ3} & RFQ length [cm] & $166.5253$ & $163.56$ ($\mathit{1.8\%}$)\\
\texttt{OBJ4} & Longitudinal emittance [MeV deg] & $0.0374$ & $0.04311$ ($\mathit{13\%}$) \\
\texttt{OBJ5} & $x$-emittance [cm mrad] & $0.0366$ & $0.03369$ ($\mathit{8.6\%}$) \\
\texttt{OBJ6} & $y$-emittance [cm mrad] & $0.0305$ & $0.02792$ ($\mathit{9.2\%}$) \\
\hline
\end{tabular}}
 \caption{\label{tab:optimal-design} Optimal design configuration and corresponding beam summary statistics of RFQ for IsoDAR's use case. This RFQ's design was created by running a multi-objective Bayesian optimization using the best-performing neural network, RFQNet2, as a surrogate model for the beam dynamics simulation. For the optimal set of design variables shown, RFQNet2's predictions for the beam summary parameters are shown alongside PARMTEQM's simulation results.}
\end{table}

%% file: Sec4_Discussion.tex
\section{Discussion}
\label{sec:discussion}

In this work, we, for the first time, build a surrogate model (RFQNet2) for an RFQ-throughgoing beam simulator that can achieve $<6\%$ MAPE on predictions of the outgoing beam dynamics. The performance of this surrogate model by objective studied can be seen in Tab.~\ref{tab:mapes}. Significant improvements in prediction accuracy were achieved by training a series of neural networks with an unusually large batch size of 1024, and by unveiling a previously hidden variable: the number of RFQ cells, which is computed deterministically from the other 14 design variables. The test-set MAPEs are improved over previous work for each of the objectives studied; in the case of transverse emittances, the MAPEs are a third of those seen in Ref.~\cite{frontiers}. However, we should note that there is no one-size-fits-all NN that achieves the lowest possible MAPE for all objectives studied; this is emphasized by the boldfaced MAPEs shown in the tables in \ref{sec:mapebyobjectiveappendix}.

\subsection{Time Costs of Surrogate Models and their Development}

The development of surrogate models, in this study, is meant to (1) reduce the computational burden that running high-fidelity simulations places on optimization routines; and (2)
facilitate quasi-real-time feedback for accelerator operators during tuning and running a machine.
The timescale of fine-tuning an RFQ can quickly balloon; a single run of PARMTEQM to compute the summary statistics of a beam traversing an RFQ of arbitrary design takes, conservatively, $30$ seconds, while the surrogate models RFQNet1 and RFQNet2 execute in just a fraction of a second.

We must point out, however, that the end-to-end development of the surrogate models presented do not, on their own, beat PARMTEQM at saving time for a single optimization. Data generation for the training and test samples took about a week of continuous running on a single workstation. Performing hyperparameter scans and cross-validation took another day of runtime. A single optimization run took about $1000$ function calls to the surrogate model.

However, considering the lengthy process of designing an
accelerator system (typically consisting of more elements than the RFQ) and iterations on design choices, it
is not realistic to assume a single optimization run will
yield the desired machine. Rather, tens of optimizations will have to be performed with ever changing output energies, beam sizes, etc. to match the RFQ to the previous and following acceleration stages.

Finally, an avenue of future study will be to understand the lower limit of feasible training sample sizes. Generation of the training data used in this study served as the main bottleneck, and a circumstance in which fewer training samples could be used to build surrogate models with comparable predictive accuracy may, eventually, prove faster than PARMTEQM.

The prototyping of an accurate surrogate model demonstrated in this study paves the way for future work using high-fidelity simulation codes based on the  
particle-in-cell (PIC) method, like  OPAL~\cite{adelmann_opal_2019} and 
WARP~\cite{vay_novel_2012}. As opposed to PARMTEQM, these codes can accurately 
calculate the behaviour of space-charge dominated beams, albeit at much higher
computation time (it can take hours to compute a beam traversing a single RFQ).
The replacement of these softwares with fast-executing surrogate models would result in orders of magnitude improvement in runtime~\cite{edelen:ml2}. Moreover, even though PARMTEQM is much faster than these higher-fidelity codes, it would still be hard to use it in a real-time commissioning tool due to its $\approx30$-second runtime, while RFQNet1 or RFQNet2 are, by comparison, virtually instantaneous. One can envision a program that allows an accelerator scientist to make small adjustments to certain controllable RFQ parameters (such as the target energy or the electromagnetic fields responsible for beam focusing, shaping, and bunching) and see, in almost real-time, a description of the outgoing beam. Such a program could find use during the running of an accelerator, where rapid computations are necessary to ensure the successful operation of the RFQ, and would further justify the need for the development of surrogate models similar to those presented here.

\subsection{Recommendations}
The significant improvements in MAPE for top-performing NNs in this work over \cite{frontiers} lead us recommend the following avenues for future work:

\begin{enumerate}
    \item \textit{Explore different neural network architectures for different objectives.} Different objective variables preferred different NN architectures (indicated explicitly in the out-of-sample MAPEs for each objective in \ref{sec:mapebyobjectiveappendix}). We can feasibly run hyperparameter scans for NNs aiming to predict one (or perhaps some subset of) objective(s) for additional boosts in performance. We are especially interested in implementing such models for the poorest-performing objective, namely, longitudinal emittance.

    \item \textit{Limit datasets to physically realistic scenarios.} A large boost in performance of the neural network was due in part to the rejection of data samples with simulated beam transmission below $60\%$. In the real-world commissioning of an RFQ, beam transmissions below $60\%$ are unrealistic, and we expect that the quality of the simulation of softwares like PARMTEQM would deteriorate in these circumstances. While the design variable boundaries were set as large as practically physically applicable when generating the initial training dataset to cover a possible large number of (even unreasonable) RFQ configurations, ensuring physically realistic scenarios enhances general data quality and predictive accuracy.
    
    \item \textit{Explore surrogate model architectures beyond fully-connected neural networks and summary statistic inputs.} We emphasize that multivariate regressive tasks are not limited to just fully-connected neural networks, or that the only way to develop an RFQ surrogate model is by feeding the exact same 14 design variables that we reference in this study. We are especially interested in the use of Residual Neural Networks (ResNets) for improving the predictive accuracy of the general multivariate regression task, as well as the use of Convolutional Neural Networks (CNNs) that can be trained on pixelated ``pictures" of the throughgoing beam, giving a more granular picture of the underlying physics. Recurrent neural networks (RNNs) may also be useful in propagating a beam iteratively through some number of RFQ cells.
\end{enumerate}

%% file: Sec5_Conclusion.tex
\section{Conclusion}
\label{sec:conclusion}
We presented an in-depth investigation of surrogate modeling for RFQ linear particle accelerators. We especially scrutinized input data preparation, and unearthed hidden variables to address the predictive shortcomings of previous works.

Correlations between the design variables did not impact the quality of the model, though we still recommend decorrelation especially when performing Bayesian optimizations necessitating complex relationships between feature variables. 

Choosing physically sensible limits and removal of nonphysical data points is essential. For example, we omitted all data with RFQ transmission $<60$\,\% from NN training, as these data points are undesirable (and sometimes even nonphysical) in the context of particle accelerator engineering.

In order to resolve asymmetries between response variables that are otherwise symmetric, such as the two transverse emittances, it is essential to carefully prepare data to include some variable having this distinguishing power. For example, determining whether the emittance in the $x$ or $y$ directions were higher was only possible if we included whether the number of RFQ cells was odd or even.

With these considerations in mind, our best-performing model was a depth 6, width 100 neural network trained on a $\approx$160,000-subsample of $>$200,000 point dataset having 14 design variables, one additional deterministic feature, and 6 objectives, which performs exceptionally well compared to previous efforts of a similar kind (we reproduce all relevant objectives with a MAPE of $<6$\,\%). To demonstrate the utility of this surrogate model, we performed a multi-objective optimization of the IsoDAR RFQ, and confirmed the
optimized design using PARMTEQM (the simulation
software used to generate the training data).

Our next steps are to test network architectures other than deep fully connected networks and to build surrogate models for other types of particle accelerators such as the IsoDAR cyclotron. These are important steps forward to creating a system of surrogate models that can speed up the development and real-time tuning of a multitude of
particle accelerator experiments.

%% file: appendix.tex
\newpage
\section{Histograms of the transformed design variable}
\label{sec:transformedhist}

\begin{figure*}[htp!]
\centering
\includegraphics[width=1.0\textwidth]{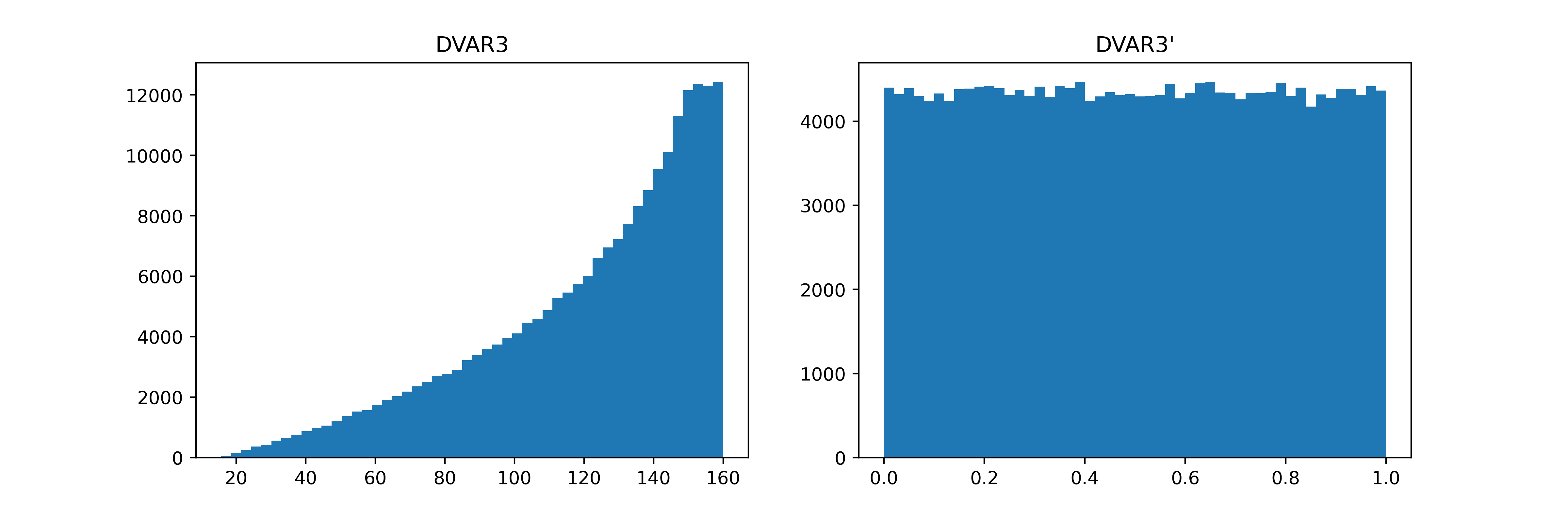}
\caption{Histograms of the unscaled (left) and transformed (right) \texttt{DVAR3}. The transformation is outlined in equation \ref{eqn:dvar3transform}.}
    \label{fig:dvar3transform}
\end{figure*}

\section{Joint distributions of true and predicted $x$- and $y$-emittances for early neural networks}
\label{sec:emittancejointdists}

In early versions of this work, we had trained a series of neural networks for which we did not specify the parity of the number of RFQ cells as a fifteenth design variable. The predicted RMS transverse emittances of the best-performing of these neural networks are shown as a joint distribution to compare to the true underlying joint distribution.

\begin{figure*}[htp!]
    \centering
    \includegraphics[width=0.8\textwidth]{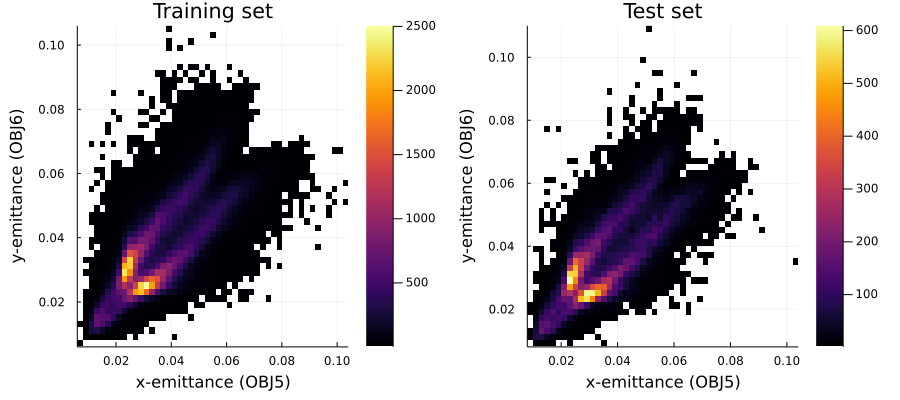}
    \caption{Joint distributions of \emph{true} $x$ and $y$ emittances (\texttt{OBJ5} and \texttt{OBJ6}, respectively) for sample training and test sets. 
    From the inherent $x$, $y$ symmetry of RFQs, we expect approximate symmetry about the $\epsilon_y = \epsilon_x$ line.}
    \label{fig:trueemits}
\end{figure*}

\begin{figure*}[htp!]
    \centering
    \includegraphics[width=0.8\textwidth]{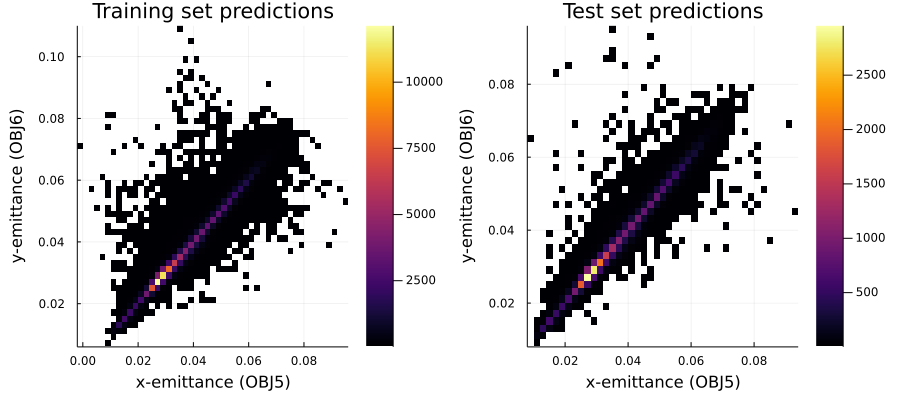}
    \caption{Joint distributions of \emph{predicted} $x$ and $y$ emittances (\texttt{OBJ5} and \texttt{OBJ6}, respectively) for sample training and test sets. Evident in these plots is the fact that the ``double band" structure discussed in Fig. \ref{fig:trueemits} is not recovered. 
    These predictions were generated by a neural network that did not include the number of cells as a fifteenth design variable.}
    \label{fig:predemits}
\end{figure*}

\newpage
\section{Out-of-sample MAPEs for scanned neural networks}
\label{sec:mapebyobjectiveappendix}

Since the validation-set $R^2$ of each of the 9 NN architectures presented in Tab.~\ref{tab:valr2} do not account for the predictive accuracies of each of the 6 objective variables separately, we report the MAPE for each objective for each neural network archtecture scanned, as the cross-fold average $\pm$ standard deviation. No one neural network observed the lowest validation set MAPE for all objective variables, highlighting the conclusion that different neural network architectures may be better equipped to predict different objectives.

\include{mapes_by_obj_notransmissioncut}
\include{mapes_by_obj_transmissioncut}


%% file: mapes_by_obj_notransmissioncut.tex
\begin{table*}[htp!]
\centering
\resizebox{\textwidth}{!}{%
\begin{tabular}{|c||c|c c c|}
\hline
\multirow{4}{*}{\rotatebox[origin=c]{90}{\texttt{OBJ1}}} & & Depth 4 & Depth 5 & Depth 6 \\ \cline{2-5}

& Width 50 & $0.01968\pm 0.0006$ & $0.01648\pm 0.000375$ & $0.01501\pm 0.000192$ \\
& Width 75 & $0.01994\pm 0.00101$ & $0.01623\pm 0.000253$ & $\mathbf{0.01473\pm 0.000238}$ \\
& Width 100 & $0.01924\pm 0.000953$ & $0.01591\pm 0.000533$ & $0.01480\pm 0.000189$ \\
    
\hline
\multicolumn{5}{c}{} \\
        
\hline
\multirow{4}{*}{\rotatebox[origin=c]{90}{\texttt{OBJ2}}} & & Depth 4 & Depth 5 & Depth 6 \\ \cline{2-5}

& Width 50 & $0.01808\pm 0.0000398$ & $0.01816\pm 0.000112$ & $0.01815\pm 0.000124$ \\
& Width 75 & $0.01809\pm 0.0000569$ & $0.01805\pm 0.0000409$ & $0.01815\pm 0.0000762$ \\
& Width 100 & $0.01806\pm 0.0000437$ & $\mathbf{0.01804\pm 0.00003}$ & $0.01818\pm 0.0000342$ \\
    
\hline
\multicolumn{5}{c}{} \\
        
\hline
\multirow{4}{*}{\rotatebox[origin=c]{90}{\texttt{OBJ3}}} & & Depth 4 & Depth 5 & Depth 6 \\ \cline{2-5}

& Width 50 & $0.01501\pm 0.000276$ & $0.01369\pm 0.000502$ & $\mathbf{0.01303\pm 0.000279}$ \\
& Width 75 & $0.01491\pm 0.0004$ & $0.01352\pm 0.000201$ & $0.01310\pm 0.000322$ \\
& Width 100 & $0.01503\pm 0.000326$ & $0.01354\pm 0.000261$ & $0.01369\pm 0.000908$ \\
    
\hline
\multicolumn{5}{c}{} \\
        
\hline
\multirow{4}{*}{\rotatebox[origin=c]{90}{\texttt{OBJ4}}} & & Depth 4 & Depth 5 & Depth 6 \\ \cline{2-5}

& Width 50 & $0.08992\pm 0.00374$ & $0.07754\pm 0.00224$ & $0.07139\pm 0.00125$ \\
& Width 75 & $0.09217\pm 0.00445$ & $0.07576\pm 0.00161$ & $\mathbf{0.07024\pm 0.00152}$ \\
& Width 100 & $0.08982\pm 0.00356$ & $0.07429\pm 0.00111$ & $0.07067\pm 0.000931$ \\
    
\hline
\multicolumn{5}{c}{} \\
        
\hline
\multirow{4}{*}{\rotatebox[origin=c]{90}{\texttt{OBJ5}}} & & Depth 4 & Depth 5 & Depth 6 \\ \cline{2-5}

& Width 50 & $0.06242\pm 0.00151$ & $0.05353\pm 0.00073$ & $0.04911\pm 0.000544$ \\
& Width 75 & $0.06092\pm 0.00173$ & $0.05089\pm 0.0011$ & $0.04713\pm 0.000374$ \\
& Width 100 & $0.05806\pm 0.00127$ & $0.04886\pm 0.00157$ & $\mathbf{0.04631\pm 0.000919}$ \\
    
\hline
\multicolumn{5}{c}{} \\
        
\hline
\multirow{4}{*}{\rotatebox[origin=c]{90}{\texttt{OBJ6}}} & & Depth 4 & Depth 5 & Depth 6 \\ \cline{2-5}

& Width 50 & $0.06194\pm 0.00173$ & $0.05338\pm 0.00126$ & $0.04900\pm 0.000365$ \\
& Width 75 & $0.06041\pm 0.00158$ & $0.05090\pm 0.00136$ & $0.04721\pm 0.000844$ \\
& Width 100 & $0.05740\pm 0.0017$ & $0.04820\pm 0.000718$ & $\mathbf{0.04637\pm 0.00125}$ \\
    
\hline
\end{tabular}}
\caption{\label{tab:notransmission-cut-mapes-by-obj} Cross-fold average $\pm$ standard deviation validation-set MAPEs by objective variable for each of the 9 NN architectures explored in the initial hyperparameter scan (Tab.~\ref{tab:hyperparamscanvalues}), trained on all available training data. RFQNet1 was selected to be the depth 5, width 100 neural network based on scores reported in Tab.~\ref{tab:valr2}, top, though this model was not the best-performing for each of the 6 objectives studied. The lowest MAPEs for each objective are shown in bold.}
\end{table*}

%% file: mapes_by_obj_transmissioncut.tex
\begin{table*}[htp!]
\centering
\resizebox{\textwidth}{!}{%
\begin{tabular}{|c||c|c c c|}

\hline
\multirow{4}{*}{\rotatebox[origin=c]{90}{\texttt{OBJ1}}} & & Depth 4 & Depth 5 & Depth 6 \\ \cline{2-5}

& Width 50 & $0.01212\pm 0.000647$ & $0.0107\pm 0.000246$ & $\mathbf{0.009651\pm 0.000243}$ \\
& Width 75 & $0.01250\pm 0.000234$ & $0.01054\pm 0.000316$ & $0.009738\pm 0.000283$ \\
& Width 100 & $0.01238\pm 0.000222$ & $0.01035\pm 0.000462$ & $0.009979\pm 0.000444$ \\
    
\hline
\multicolumn{5}{c}{} \\
        
\hline
\multirow{4}{*}{\rotatebox[origin=c]{90}{\texttt{OBJ2}}} & & Depth 4 & Depth 5 & Depth 6 \\ \cline{2-5}

& Width 50 & $0.01801\pm 0.000287$ & $\mathbf{0.01786\pm 0.000076}$ & $0.01793\pm 0.0000767$ \\
& Width 75 & $0.01787\pm 0.0000562$ & $0.01792\pm 0.000131$ & $0.01801\pm 0.0000828$ \\
& Width 100 & $0.01799\pm 0.00017$ & $0.01793\pm 0.000221$ & $0.01797\pm 0.0000958$ \\
    
\hline
\multicolumn{5}{c}{} \\
        
\hline
\multirow{4}{*}{\rotatebox[origin=c]{90}{\texttt{OBJ3}}} & & Depth 4 & Depth 5 & Depth 6 \\ \cline{2-5}

& Width 50 & $0.01479\pm 0.00132$ & $0.01304\pm 0.00027$ & $\mathbf{0.01227\pm 0.000135}$ \\
& Width 75 & $0.01385\pm 0.000213$ & $0.01291\pm 0.00032$ & $0.01270\pm 0.000252$ \\
& Width 100 & $0.01467\pm 0.000487$ & $0.01295\pm 0.000215$ & $0.01289\pm 0.000336$ \\
    
\hline
\multicolumn{5}{c}{} \\
        
\hline
\multirow{4}{*}{\rotatebox[origin=c]{90}{\texttt{OBJ4}}} & & Depth 4 & Depth 5 & Depth 6 \\ \cline{2-5}

& Width 50 & $0.06922\pm 0.0017$ & $0.06335\pm 0.00356$ & $0.05802\pm 0.00209$ \\
& Width 75 & $0.07061\pm 0.000696$ & $0.06256\pm 0.00365$ & $\mathbf{0.05704\pm 0.000965}$ \\
& Width 100 & $0.06896\pm 0.0014$ & $0.0598\pm 0.0014$ & $0.05864\pm 0.00353$ \\
    
\hline
\multicolumn{5}{c}{} \\
        
\hline
\multirow{4}{*}{\rotatebox[origin=c]{90}{\texttt{OBJ5}}} & & Depth 4 & Depth 5 & Depth 6 \\ \cline{2-5}

& Width 50 & $0.05445\pm 0.00153$ & $0.04895\pm 0.00241$ & $0.04367\pm 0.000593$ \\
& Width 75 & $0.05528\pm 0.0027$ & $0.04704\pm 0.00243$ & $0.04236\pm 0.000757$ \\
& Width 100 & $0.05177\pm 0.00119$ & $0.04313\pm 0.000659$ & $\mathbf{0.04110\pm 0.000769}$ \\
    
\hline
\multicolumn{5}{c}{} \\
        
\hline
\multirow{4}{*}{\rotatebox[origin=c]{90}{\texttt{OBJ6}}} & & Depth 4 & Depth 5 & Depth 6 \\ \cline{2-5}

& Width 50 & $0.05438\pm 0.00151$ & $0.04818\pm 0.00143$ & $0.04310\pm 0.000202$ \\
& Width 75 & $0.05398\pm 0.00156$ & $0.04596\pm 0.000912$ & $0.04257\pm 0.00107$ \\
& Width 100 & $0.05159\pm 0.00109$ & $0.04305\pm 0.000586$ & $\mathbf{0.04095\pm 0.000797}$ \\
    
\hline
\end{tabular}}
\caption{\label{tab:transmission-cut-mapes-by-obj} Cross-fold average $\pm$ standard deviation validation-set MAPEs by objective variable, for each of the 9 NN architectures explored in the initial hyperparameter scan (Tab.~\ref{tab:hyperparamscanvalues}), trained on samples having transmissions $\geq 60\%$. RFQNet2 was selected to be the depth 6, width 100 neural network based on scores reported in Tab.~\ref{tab:valr2}, bottom, though different objectives preferred different NN architectures. The lowest MAPEs for each objective are shown in bold.}
\end{table*}